\documentclass[structabstract]{aa}
\usepackage{graphicx}
\usepackage[bf]{caption}
\usepackage[lofdepth,lotdepth]{subfig}
\usepackage{txfonts}
\usepackage{dcolumn}
\usepackage{natbib}
\usepackage{amstext}

\begin{document}

\title{Wisps in the Galactic center: NIR triggered observations of the radio source Sgr\,A* at 43\,GHz}
\titlerunning{Sgr\,A*: NIR triggered observations at 43\,GHz}
\author{C. Rauch \inst{1}
       \and
       E. Ros \inst{1,2,3}
       \and
       T. P. Krichbaum \inst{1}
       \and 
       A. Eckart \inst{1,4}
       \and
       J. A. Zensus \inst{1}
       \and
       B. Shahzamanian \inst{4}
       \and
       K. Mu\v{z}i\'{c} \inst{5,6}
       }

\institute{Max-Planck-Institut f\"ur Radioastronomie, Auf dem H\"ugel 69, 53121 Bonn, Germany\\
           \email{crauch, tkrichbaum, ros, azensus@mpifr-bonn.mpg.de}
           \and
           Observatori Astr\`onomic, Universitat de Val\`encia, E-46071, Spain
           \and
           Departament d'Astronomia i Astrof\'isica, Universitat de Val\`encia, E-46071, Spain 
           \and
           I. Physikalisches Institut, Universit\"at zu K\"oln,
           Z\"ulpicher Str. 77, 50937 K\"oln, Germany\\
           \email{eckart, shahzamani@ph1.uni-koeln.de}
           \and Nucleo de Astronom\'ia, Facultad de Ingenier\'ia, Universidad Diego Portales, Av. Ejercito 441, Santiago, Chile
           \and European Southern Observatory, Alonso de C\'ordova 3107, Casilla 19, Santiago, 19001, Chile\\
           \email{kmuzic@eso.org}
           }

\date{Submitted 2015 / revised XXX}
\abstract{The compact radio and near-infrared (NIR) source Sagittarius A* (Sgr\,A*) associated with the supermassive black hole in the Galactic center was observed at 7\,mm in the context of a NIR triggered global Very Long Baseline Array (VLBA) campaign.}
{Sgr\,A* shows variable flux densities ranging from radio through X-rays. These variations sometimes appear in spontaneous outbursts that are referred to as flares. Multi-frequency observations of Sgr\,A* provide access to easily observable parameters that can test the currently accepted models that try to explain these intensity outbursts.}
{On May 16-18, 2012 Sgr\,A* has been observed with the VLBA at 7\,mm (43\,GHz) for 6 hours each day during a global multi-wavelength campaign. These observations were triggered by a NIR flare observed at the Very Large Telescope (VLT). Accurate flux densities and source morphologies were acquired.}
{The total 7\,mm flux of Sgr\,A* shows only minor variations during its quiescent states on a daily basis of 0.06\,Jy. An observed NIR flare on May 17 was followed $\sim$4.5\,hours later by an increase in flux density of 0.22\,Jy at 43\,GHz. This agrees
well with the expected time delay of events that are casually connected by adiabatic expansion. Shortly before the peak of the radio flare, Sgr\,A* developed a secondary radio off-core feature at 1.5\,mas toward the southeast. Even though the closure phases are too noisy to place actual constraints on this feature, a component at this scale together with a time delay of $4.5\pm0.5$\,hours between the NIR and radio flare provide evidence for an adiabatically expanding jet feature.}{}
\keywords{Galaxy: center - galaxies: individual: Sgr\,A* - techniques: interferometric - galaxies: nuclei}
\maketitle

\begin{figure*}[]
\centering
\begin{tabular}{ccc}
\centering
\subfloat[][]{
\includegraphics[width=\textwidth,angle=0, scale=0.25]{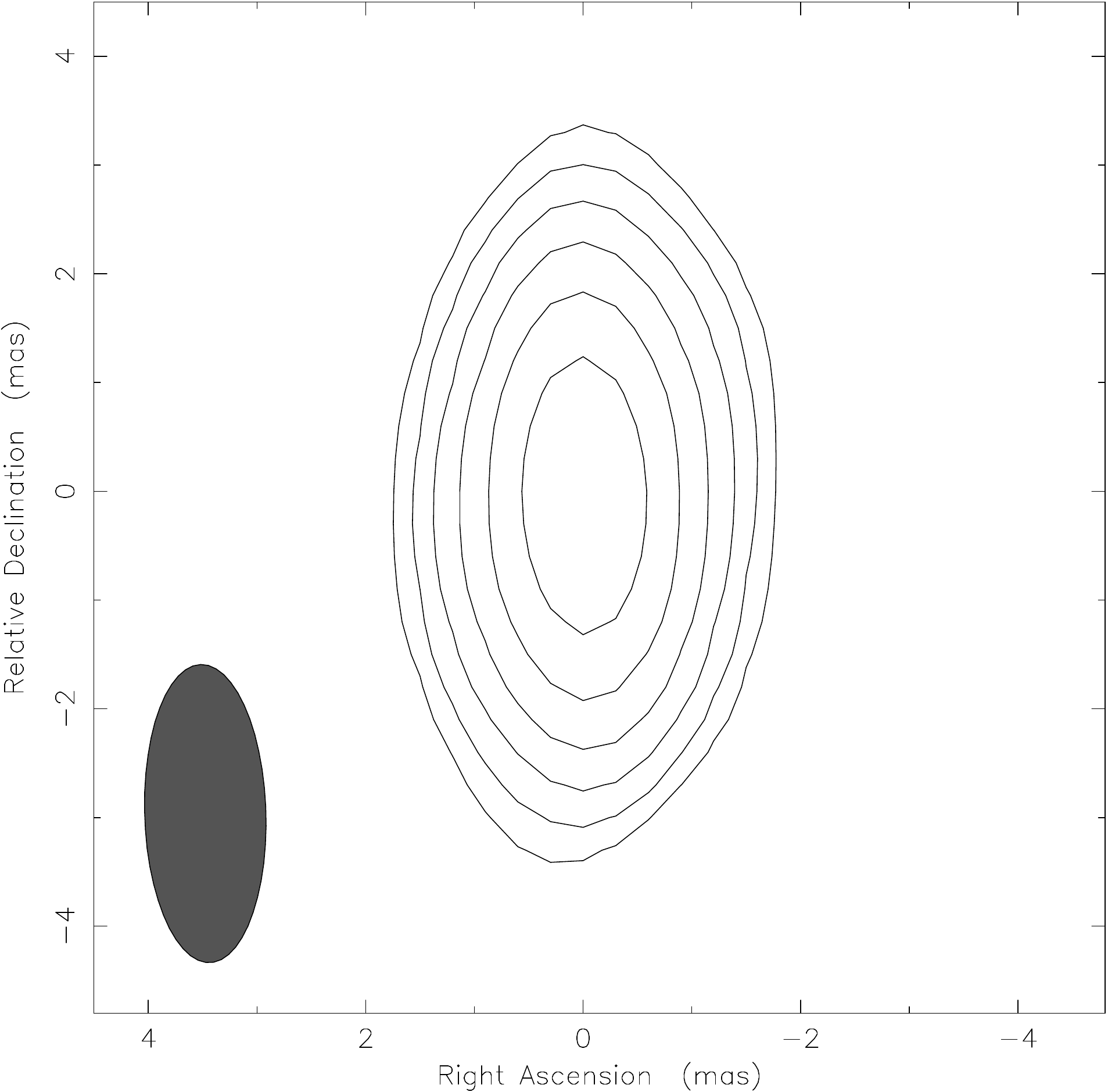}
\label{be061a}}
\qquad
\subfloat[][]{
\includegraphics[width=\textwidth,angle=0, scale=0.25]{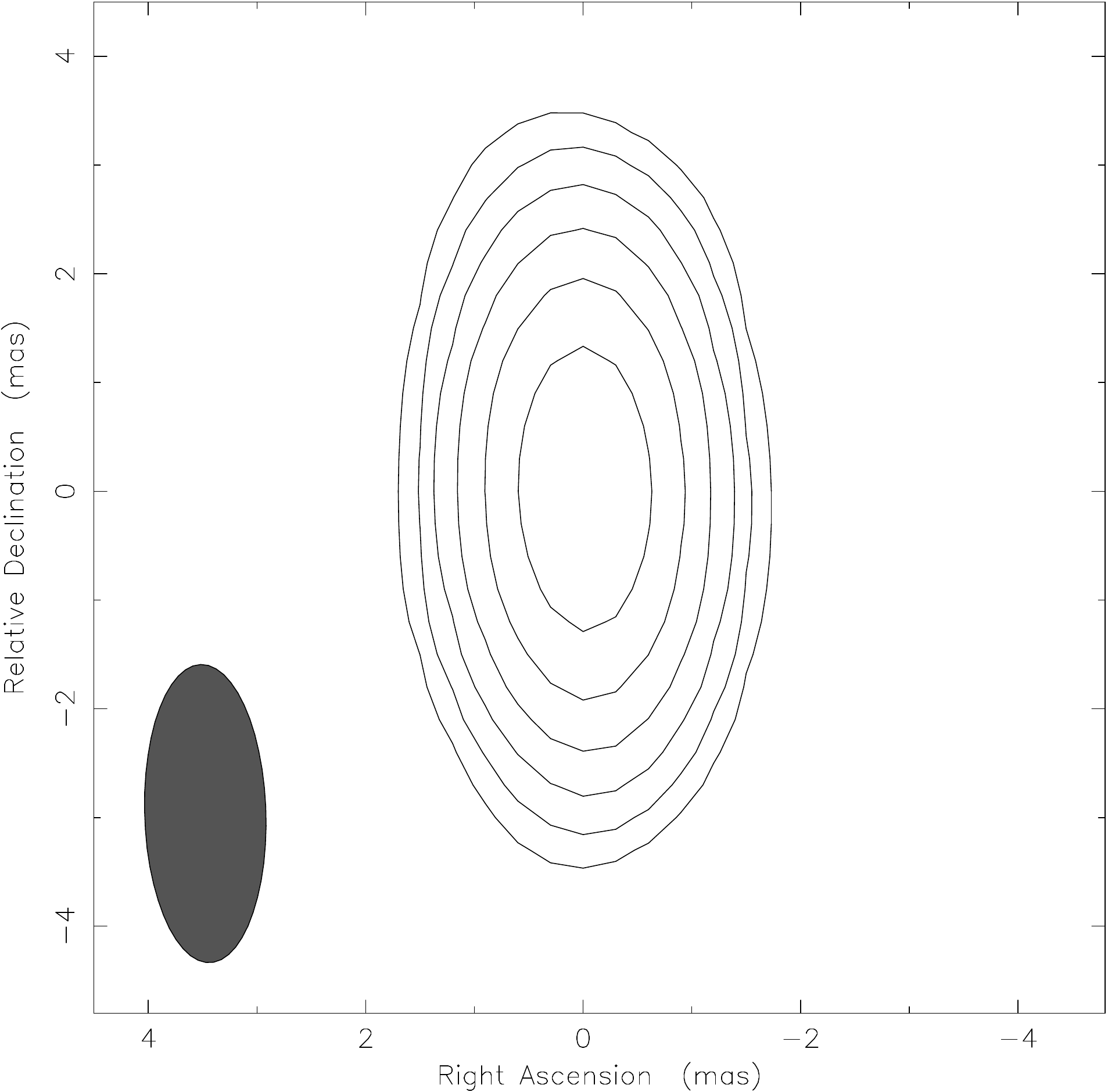}
\label{be061b}}
\qquad
\subfloat[][]{
\includegraphics[width=\textwidth,angle=0, scale=0.25]{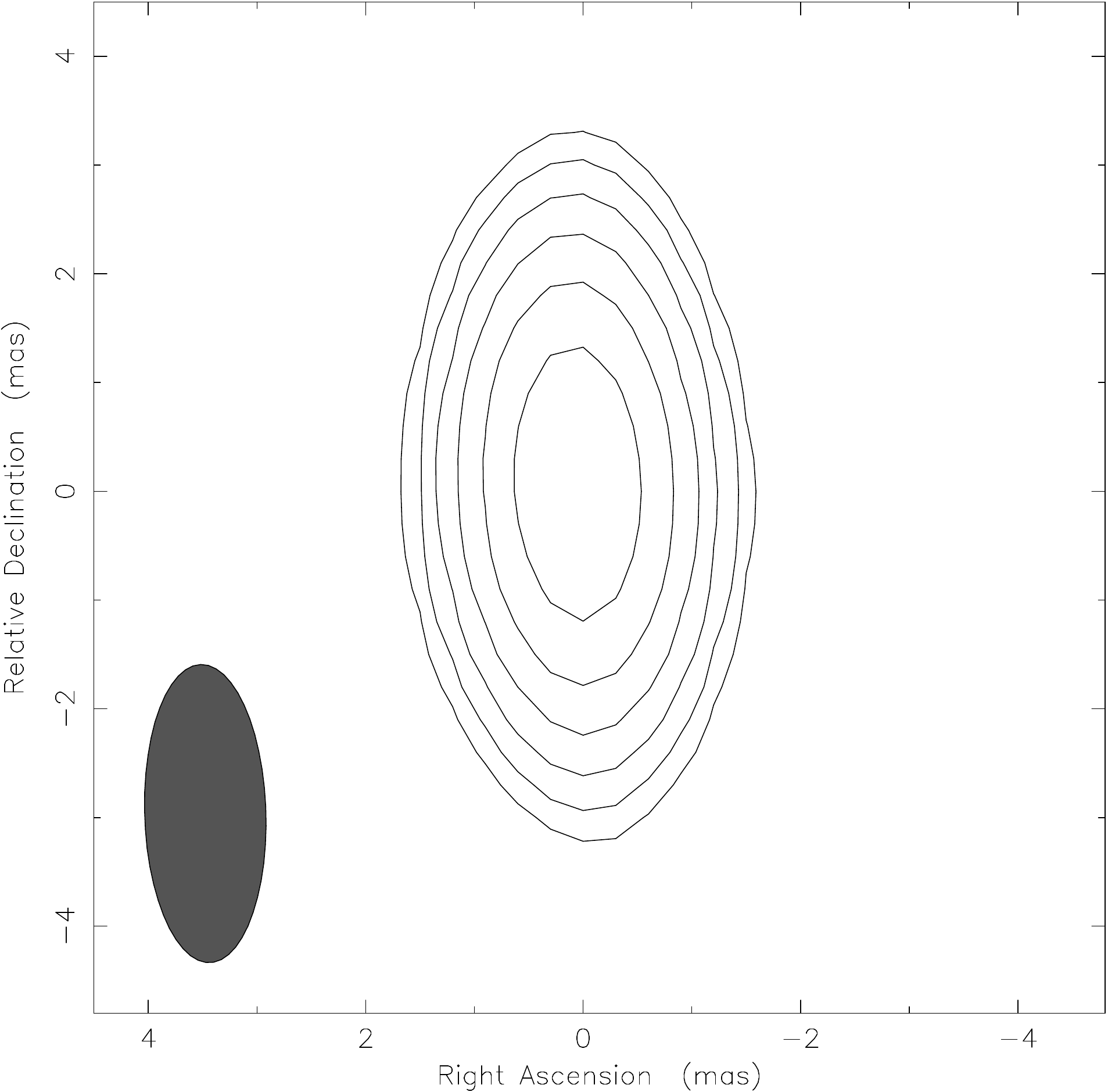}
\label{be061c}}
\end{tabular}
\centering
\caption{Uniformly weighted clean LCP maps of Sgr\,A* observed on (a) May 16 2012 (7:55 - 12:19 UT) with a flux of $1.11\pm0.02$. (b) May 17 2012 (6:04 - 12:19 UT) with a  flux of $1.34\pm0.03$. (c) May 18 2012 (7:32 - 12:19 UT) with a  flux of $1.17\pm0.02$. The maps are restored with a beam of 2.74$\times$1.12 at 1.76$^\circ$. The contour levels are 1.73\%, 3.46\%, 6.93\%, 13.9\%, 27.7\%, and 55.4\%.}
\label{be061abc}
\end{figure*}

\section{Introduction}
\label{introduction}
The compact radio source Sagittarius A* (Sgr\,A*) is commonly assumed to be associated with the supermassive black hole (SMBH) of $\sim$4.0$\times 10^6$$\,$M$_{\sun}$ at a solar distance of $\sim$8.0$\,$kpc in the center of the Milky Way (\citealt{eckart2002,schoedel2002,schoedel2003,ghez2003,ghez2008,gillessen2009}). Despite the fact that Sgr\,A* is extremely dim in terms of Eddington luminosity, based on the correlation between black hole (BH) mass and the velocity dispersion, most if not all Galactic nuclei contain an SMBH at their centers (e.g., \citealt{richstone1998,gebhardt2000,ferrarese2000}). Because it is about one hundred times closer than the second nearest Galactic nucleus (M31) and because it has the largest projected Schwarzschild radius on the sky, Sgr\,A* is the most interesting target in which to study the physics of these objects.

The intensity of Sgr\,A* shows spontaneous flux density outbursts at radio to X-ray frequencies, commonly referred to as flares (\citealt{mauerhan2005,marrone2006,yusef-zadeh2008,eckart2008a,eckart2008b,eckart2008c,eckart2009,eckart2012,lu2011,miyazaki2012}). These intensity irregularities appear on timescales ranging from 1-2 hours (main flares) down to 7-10 minutes (sub-flares) (\citealt{eckart2006a}) with stronger activity at shorter wavelengths (\citealt{baganoff2001,genzel2003,ghez2004,eckart2006b,eckart2006c}).  The short timescales suggest that this feature originates in a very compact region, possibly located close to the event horizon of the BH. Even though the detection of structures on these scales has not been achieved yet, \cite{doeleman2008,doeleman2009} and \cite{fish2011} have detected an intrinsic source structure of $\sim$40\,$\mu$as on event-horizon scales using Very Long Baseline Interferometry (VLBI) at 0.7\,mm and 1.3\,mm. 

Additionally, its apparent size is frequency dependent because of the scatter broadening of the interstellar medium (\citealt{bower2006,krichbaum2006}). It is expected that the $\lambda$$^2$-dependency breaks at around 43\,GHz and the intrinsic source size is no longer dominated by scatter broadening and thus reveales its true structure (\citealt{krichbaum1998,lo1998,doeleman2001,bower2004,shen2005,bower2006,krichbaum2006}), while \cite{bower2014} reported that strong interstellar scattering is still present at 7\,mm. Therefore, observations of Sgr\,A* with 7\,mm VLBI are of particular interest to study its source size and morphology.

The current literature provides several models that try to explain the observed variability by a relativistically aberrated accretion disk emission, hot spots, inhomogeneities in the accretion disk, or a jet (\citealt{falcke2000a,falcke2000b,broderick2006,huang2007,broderick2009}). The hot spot model proposed by \cite{yuan2009} describes in analogy to the coronal mass ejection of the Sun that magnetic flux ropes can be formed in the accretion disk by Parker instabilities. If this system looses its equilibrium stability, the material is rapidly expelled and the flux rope is propelled away from the accretion disk. Such hot spots orbiting at detectable radii would alter the morphology of Sgr\,A* at timescales corresponding to their orbit periodicity. In this case, closure phases would periodically deviate from zero. Another currently discussed model is a temporary jet anchored at Sgr\,A* (e.g., \citealt{markoff2007}). A jet can be induced by a higher accretion rate than during quiescent states or by other instabilities in the accretion flow, and temporary jets would also alter the morphology of Sgr\,A*, resulting in non-zero closure phases.

Relativistic magnetohydrodynamic simulations predict a constant size and shape of the Sgr\,A* emission region (e.g., \citealt{chan2009}). However, if adiabatic expansion were the cause of these flares, a change of the morphology and/or a full width at half-maximum
(FWHM) would be observed. An orbiting or asymmetrically located expanding feature would, on the other hand, be detectable by a position deviation of the emission center caused by the change of its structure. The associated closure phases would also deviate from from zero as a result of the increased asymmetry of the source. Therefore, the best properties to investigate the nature of the Sgr\,A* flare activity are the position, the morphology, and the FWHM Gaussian size of the compact radio source Sgr\,A*.

While the flux variability of Sgr\,A* has been studied by many authors (e.g., \citealt{zhao2003,mauerhan2005,marrone2006,marrone2008,doeleman2008,eckart2008a,eckart2008b,eckart2008c,eckart2009,eckart2012,yusef-zadeh2008,yusef-zadeh2009,li2009,kunneriath2010,sabha2010,zamaninasab2010,fish2011,miyazaki2012}), size measurements during flares have not been performed as frequently (e.g., \citealt{rogers1994,krichbaum1998,lo1998,doeleman2001,doeleman2008,shen2005,shen2006,huang2007,lu2011,bower2014}). Recently, \cite{bower2014} performed triggered multi-frequency VLBI observations of Sgr\,A* at NIR, X-ray, and 43\,GHz that revealed an elliptical intrinsic source size of $35.4 \times 12.6$\,$R_{s}$ with an rms variation of $\sim$5\% and maximum peak-to-peak change of $\lessapprox$15\%. \cite{akiyama2013} found even stronger fluctuations of its intrinsic size of 19\% at 43\,GHz based on VERA observations of the emitting region. \cite{lu2011} also reported a tendency for the minor axis to increase during higher flux periods. 
The positional change of Sgr\,A* has been investigated by \cite{reid2008}, who found an average centroid change of Sgr\,A* at 7\,mm of $71\pm45$\,$\mu$as for timescales between 50 and 100 minutes and $113\pm50$\,$\mu$as for 100-200 minutes, which rules out hot spots that contribute more than 30\% of the total flux with orbital radii above 15 $\frac{G M_\mathrm{Sgr\,A^{*}}}{c^{2}}\approx$80\,$\mu$as.  

This work presents a multi-epoch measurement campaign using 7\,mm Very Long Baseline Array (VLBA) data triggered by near-infrared (NIR) observations to investigate the accretion physics and multi-wavelength variability properties of Sgr\,A*.

\begin{figure}[]
\centering
\includegraphics[width=\textwidth,angle=0, scale=0.4]{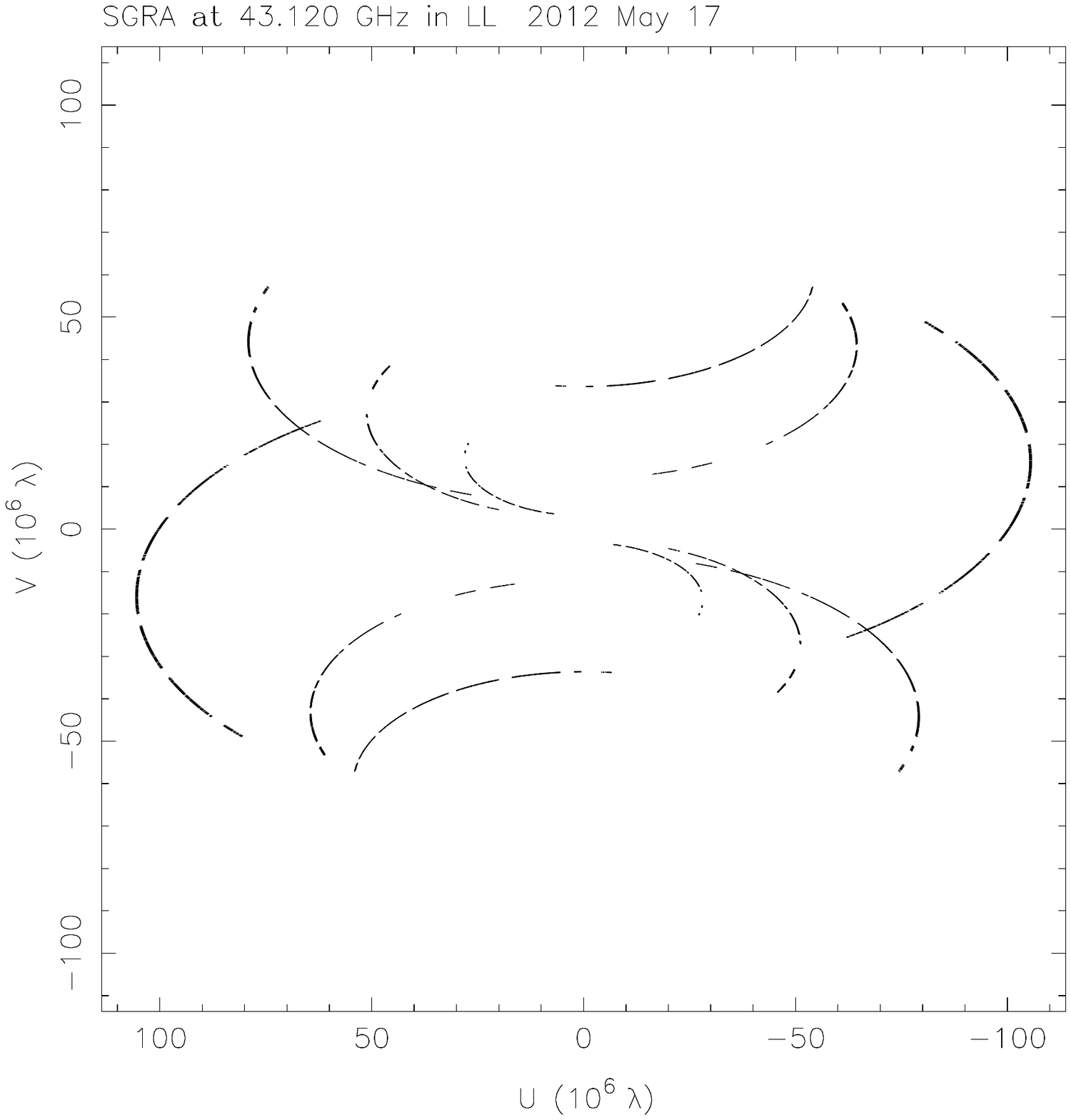}
\caption{\small UV-coverage of Sgr\,A* on May 17 2012 (6:04 - 12:19 UT) for the stations FD, KP, LA, and PT.}
\label{uvcover_sgra}
\end{figure}

\begin{figure}[]
\centering
\subfloat[][]{
\includegraphics[width=\textwidth,angle=0, scale=0.31]{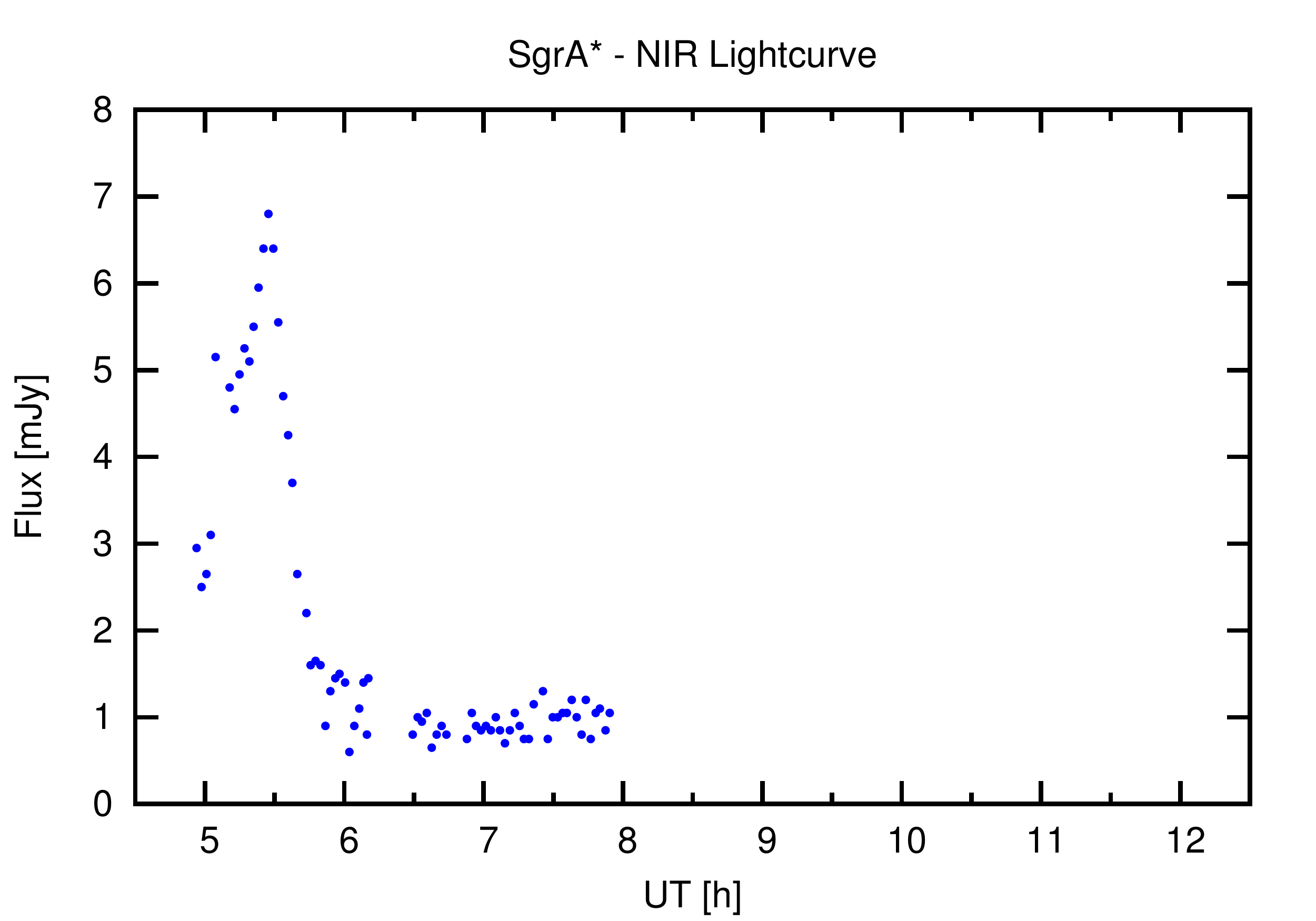}
\label{NIR_lightcurve}}
\qquad
\subfloat[][]{
\includegraphics[width=\textwidth,angle=0, scale=0.32]{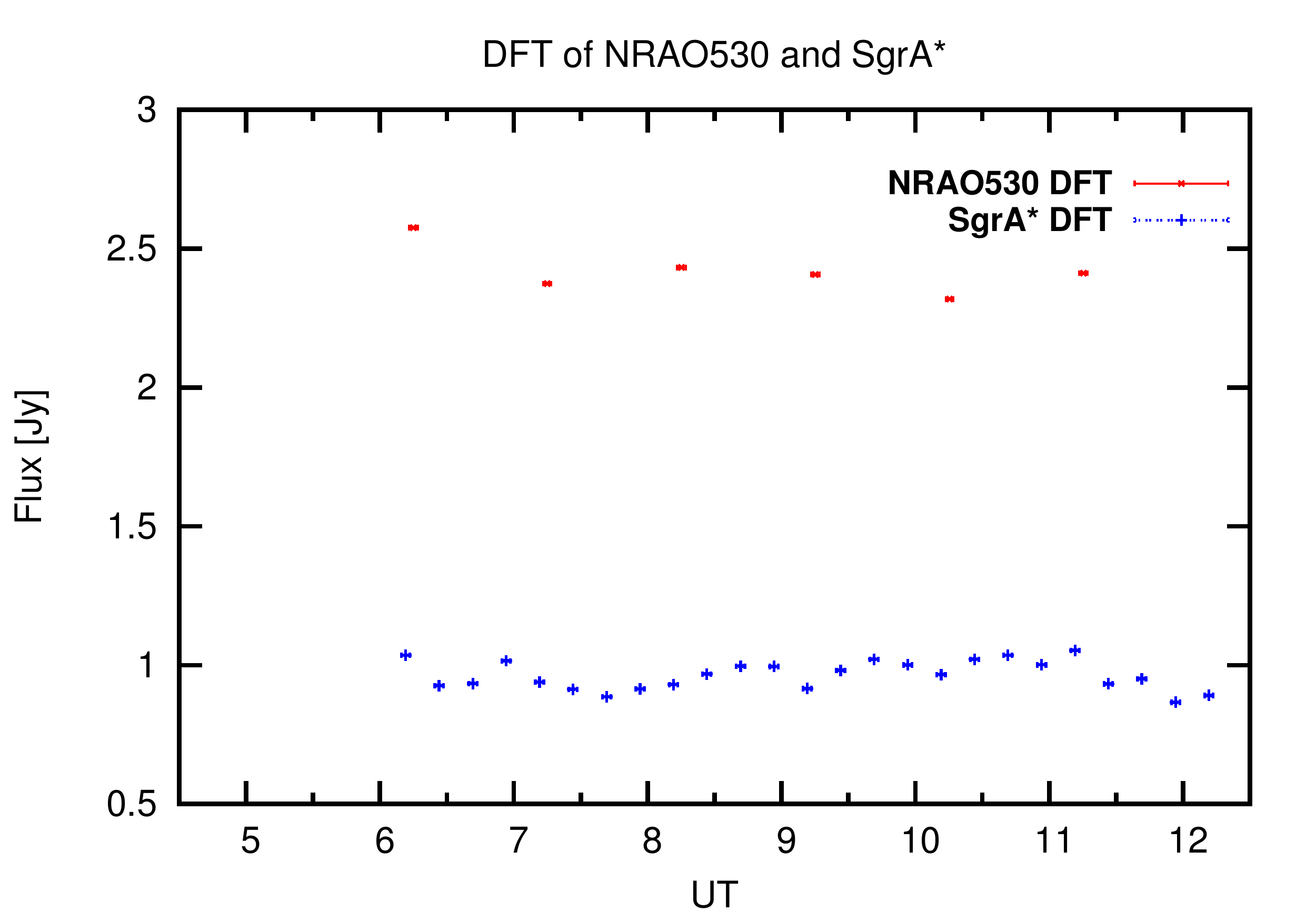}
\label{be061_sgra_nrao530_7mm_lightcurve}}
\qquad
\subfloat[][]{
\includegraphics[width=\textwidth,angle=0, scale=0.32]{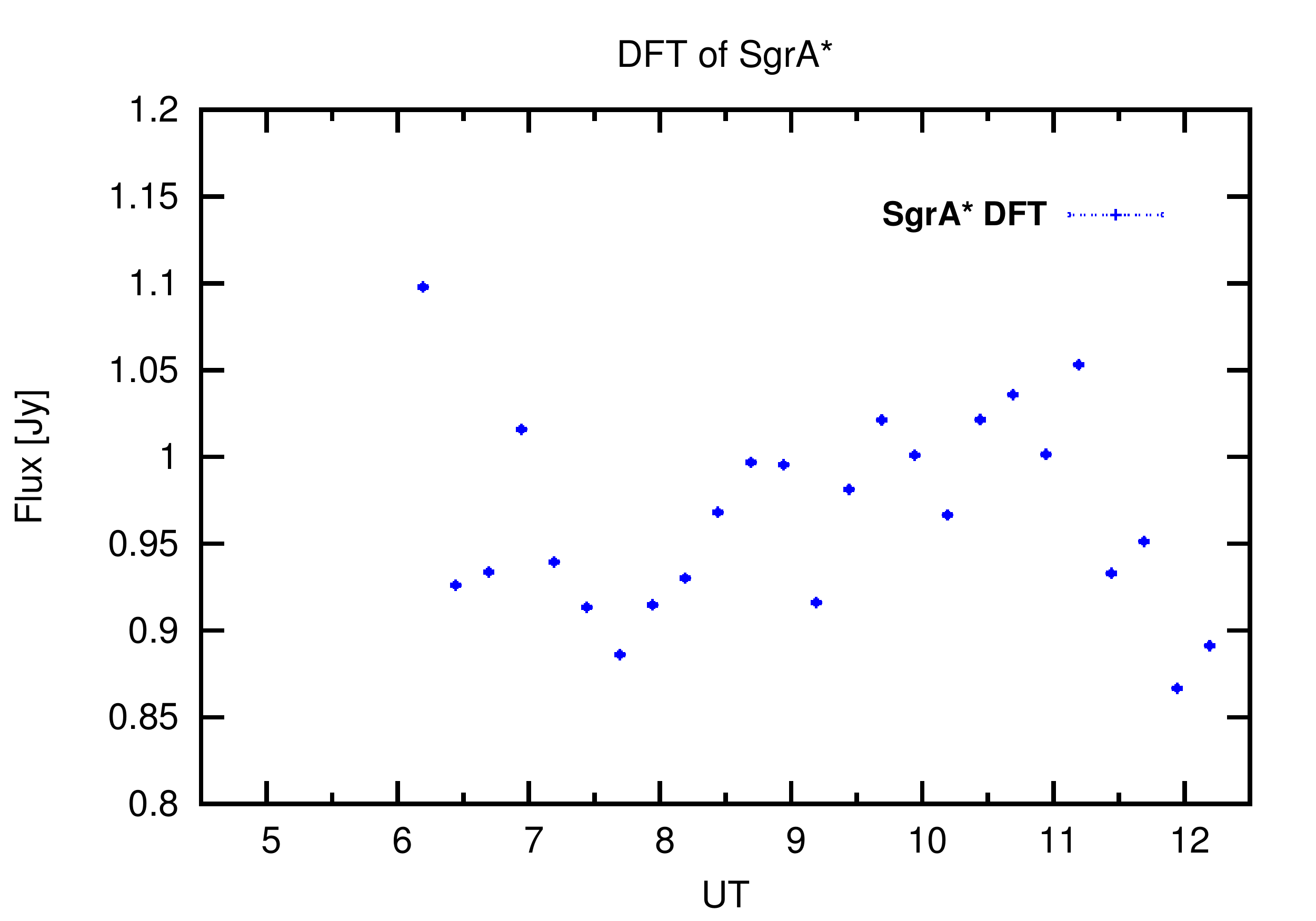}
\label{be061b_sgra_detrend}}
\qquad
\subfloat[][]{
\includegraphics[width=\textwidth,angle=0, scale=0.32]{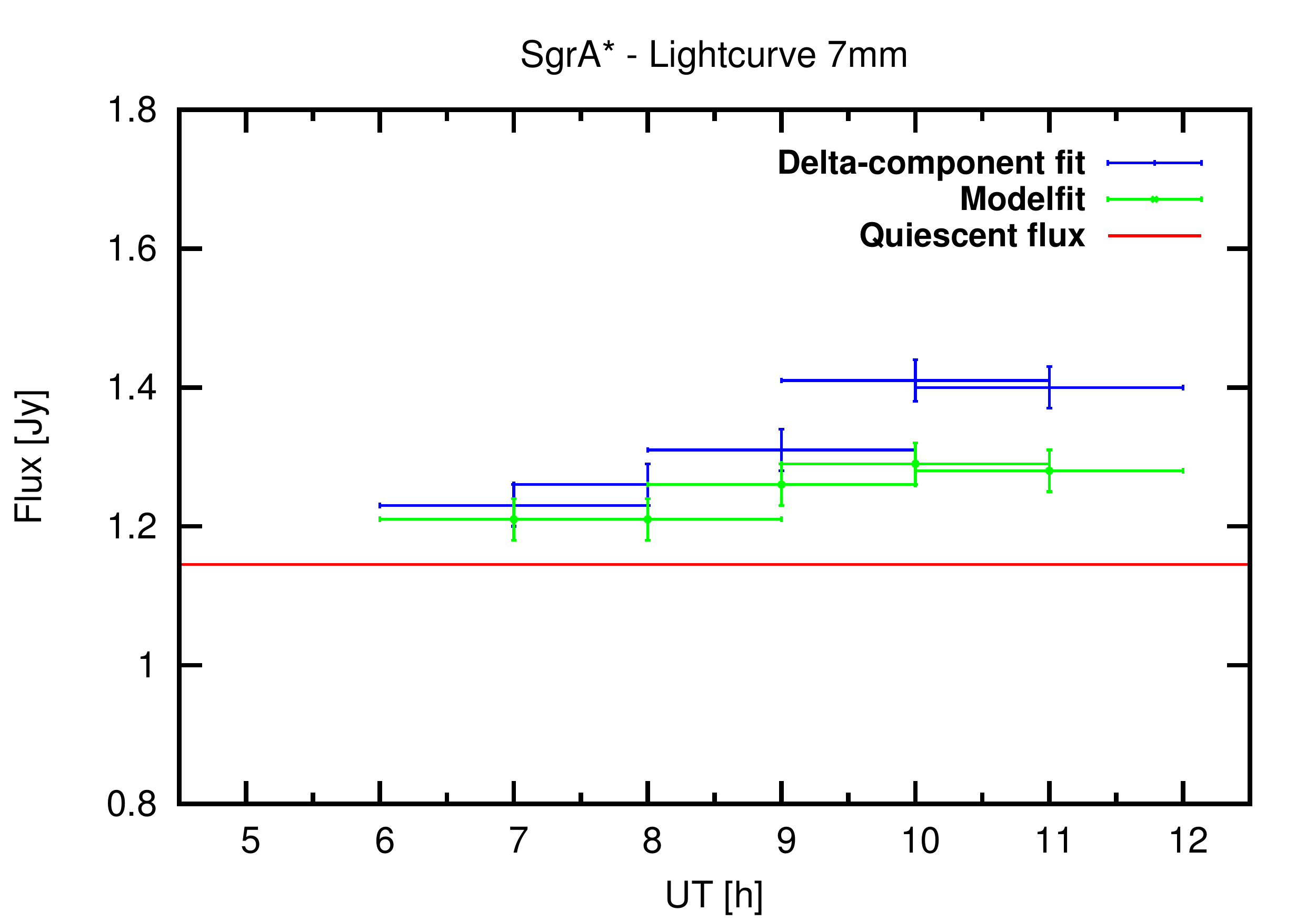}
\label{7mm_lightcurve}}
\caption{NIR and 7\,mm light curves on May 17 6:04-12:19\,h UT 2012. (a) K$_{s}$-band (2.2\,$\mu$m) light curve of Sgr A* observed in polarimetry mode on 17 May 2012. The light curve shown is produced by combining pairs of orthogonal polarization channels: 0$^\circ$ and 90$^\circ$ (for further details see \cite{shahzamanian2015}). (b) 7\,mm DFT of NRAO\,530 and Sgr\,A* on May 17. Sgr\,A* was averaged over 15\,min. NRAO\,530 was averaged over 2\,min. (c) Detrended 7\,mm DFT of Sgr\,A* on May 17. (d) 7\,mm light curve derived from the two-hour maps observed on May 17 (see Fig. \ref{sgra2hmaps}). Fluxes were obtained from delta-component maps (green) and model fitting (blue). Errors represent the formal errors of 1.7\% derived from NRAO\,530. Not shown are the systematic errors of $\sim$18\%, which should be corrected for by calibration.}
\label{lightcurve}
\end{figure}

\section{Observations and data analysis}
\label{obsdata}
\subsection{Radio and NIR data}
\label{vlbadata}
The presented 7\,mm VLBA data of Sgr\,A* were observed on three consecutive days on May 16-18, 2012 for $\sim$6 hours each day in dual polarization mode (see Fig. \ref{be061abc}, stations: Fort Davis, Hancock, Kitt Peak, Los Alamos, Mauna Kea, Owens Valley, Pie Town, Brewster and North Liberty). These observations were triggered by a preceding NIR flare observed at the Very Large Telescope (VLT). Dual circular polarization was recorded at each station at an aggregate bit rate of 2\,Gbps (8 intermediate frequency (IF) channels at 32\,MSamples/sec of 2\,bit) using the VLBA/Mk5c correlator.

The science target (Sgr\,A*) and the compact extragalactic sources (J1745$-$283, J1748$-$291) were observed in a duty cycle of 60\,s rapidly switching to Sgr\,A* between each pointing (J1745$-$283$\rightarrow$Sgr\,A*$\rightarrow$J1748$-$291$\rightarrow$Sgr\,A*$\rightarrow$J1745$-$283). The individual scan length on each source was $\sim$6\,s. 

The calibrators 3C\,345 and NRAO\,530 were observed every hour for 2\,min each; they served as fringe tracers and amplitude calibrators. The third calibrator 3C\,279 was targeted only at the beginning of each day for 2\,min between one (first and third day) and three times (second day). The data were correlated at the VLBA correlator in Socorro, NM, USA with an integration time of 1\,s at 512 spectral bands per baseband channel.

The trigger was performed on a single-day 2.2\,$\mu$m NIR dataset. These observations were performed on May 17 4:55-7:51\,UT at the VLT with the NACO instrument. Further information on NIR observations can be gained from \cite{shahzamanian2015}, who provide a detailed discussion of this dataset.

\begin{figure*}[]
\centering
\subfloat[][]{
\includegraphics[width=\textwidth,angle=0, scale=0.22]{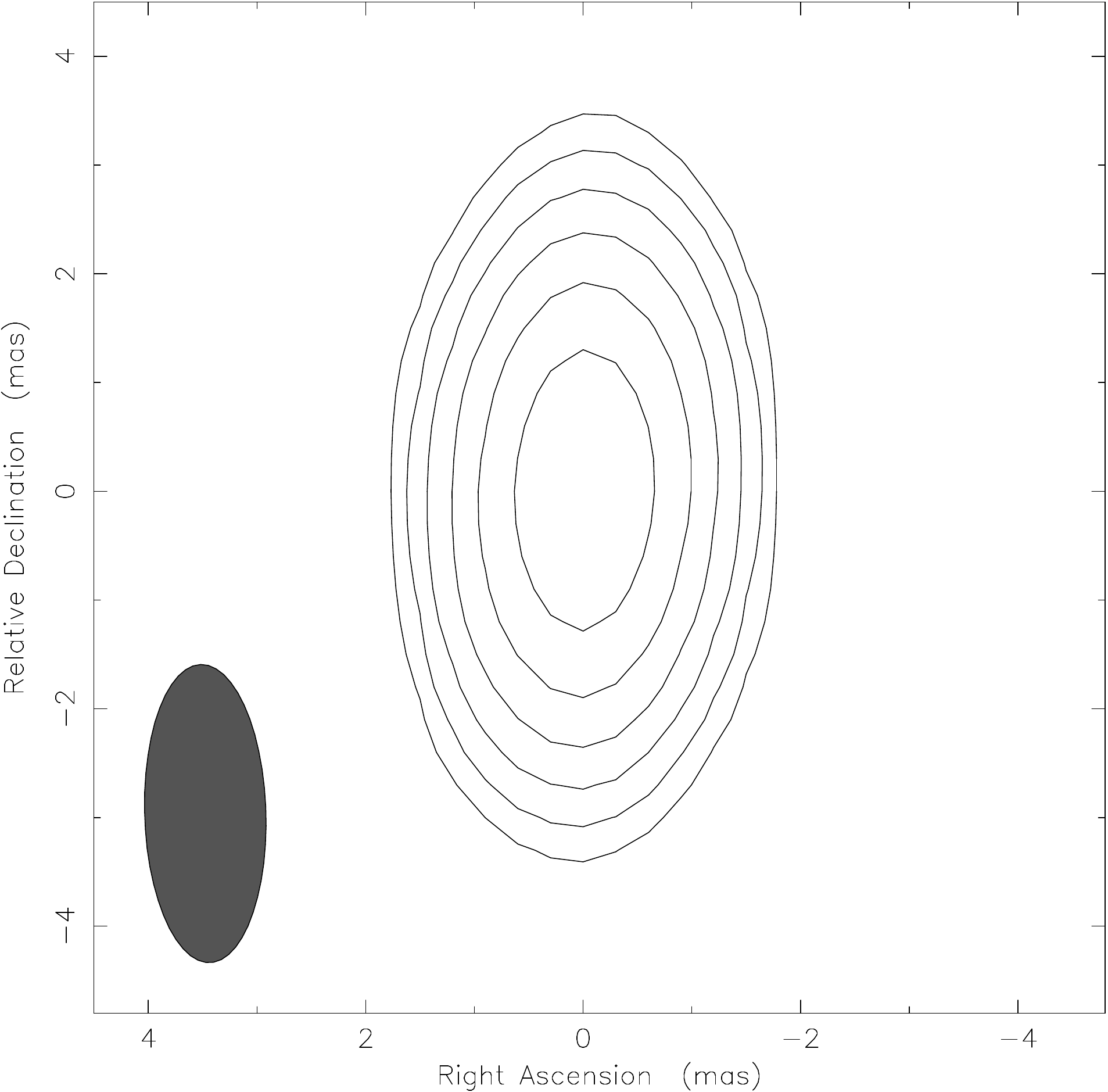}
\label{sgra6-8}}
\qquad
\subfloat[][]{
\includegraphics[width=\textwidth,angle=0, scale=0.22]{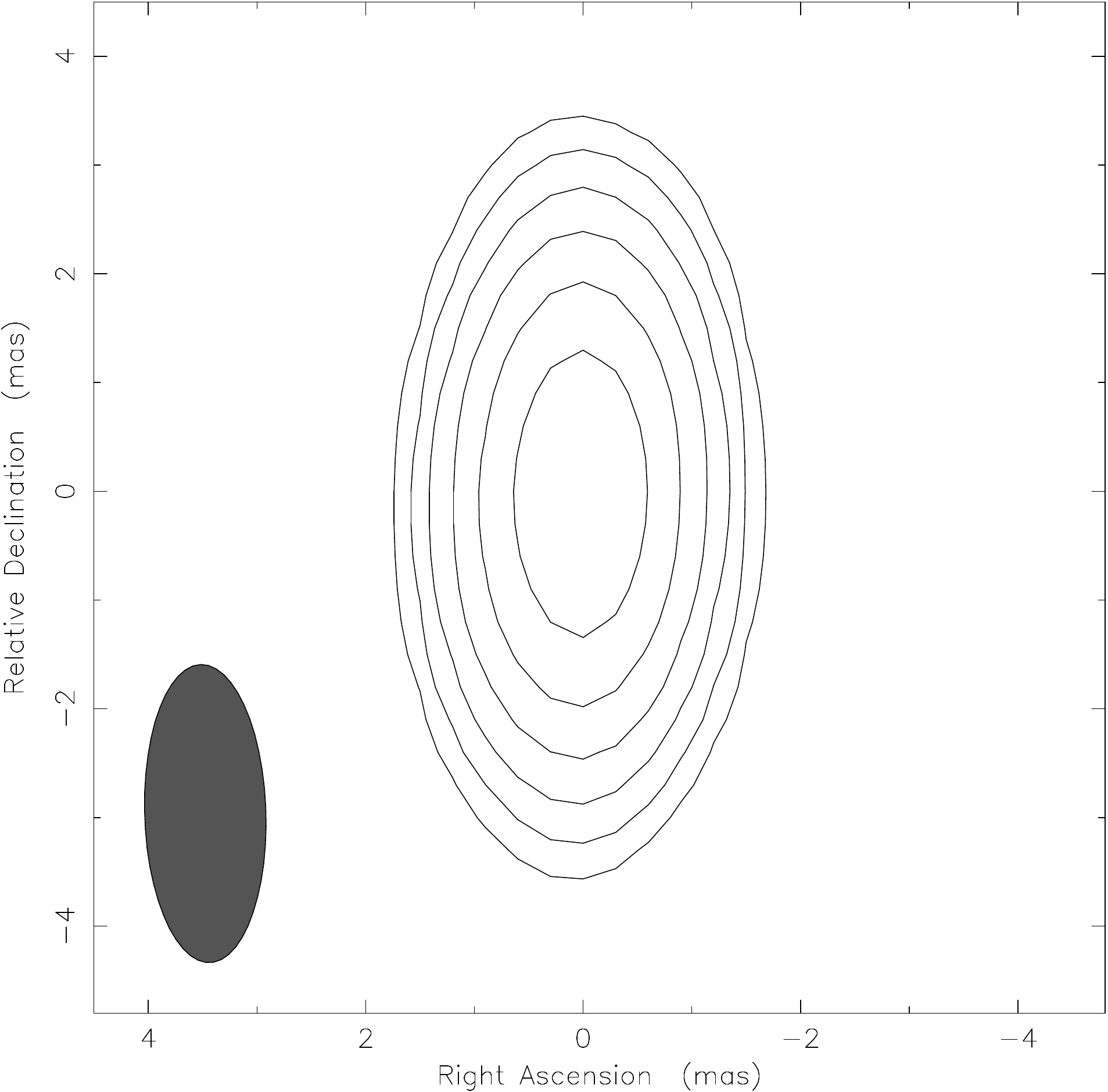}
\label{sgra7-9}}
\qquad
\subfloat[][]{
\includegraphics[width=\textwidth,angle=0, scale=0.22]{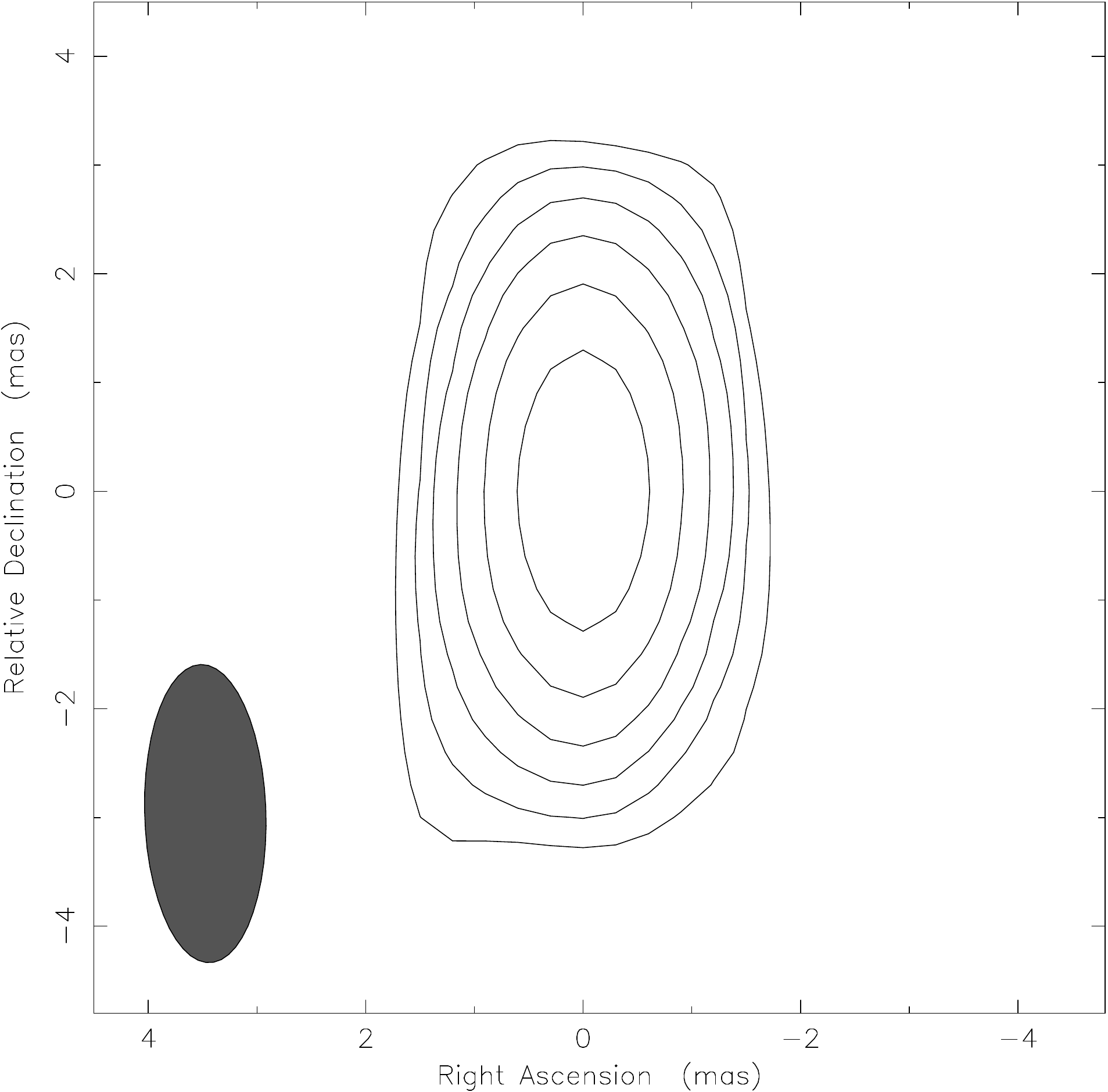}
\label{sgra8-10}}
\qquad
\subfloat[][]{
\includegraphics[width=\textwidth,angle=0, scale=0.22]{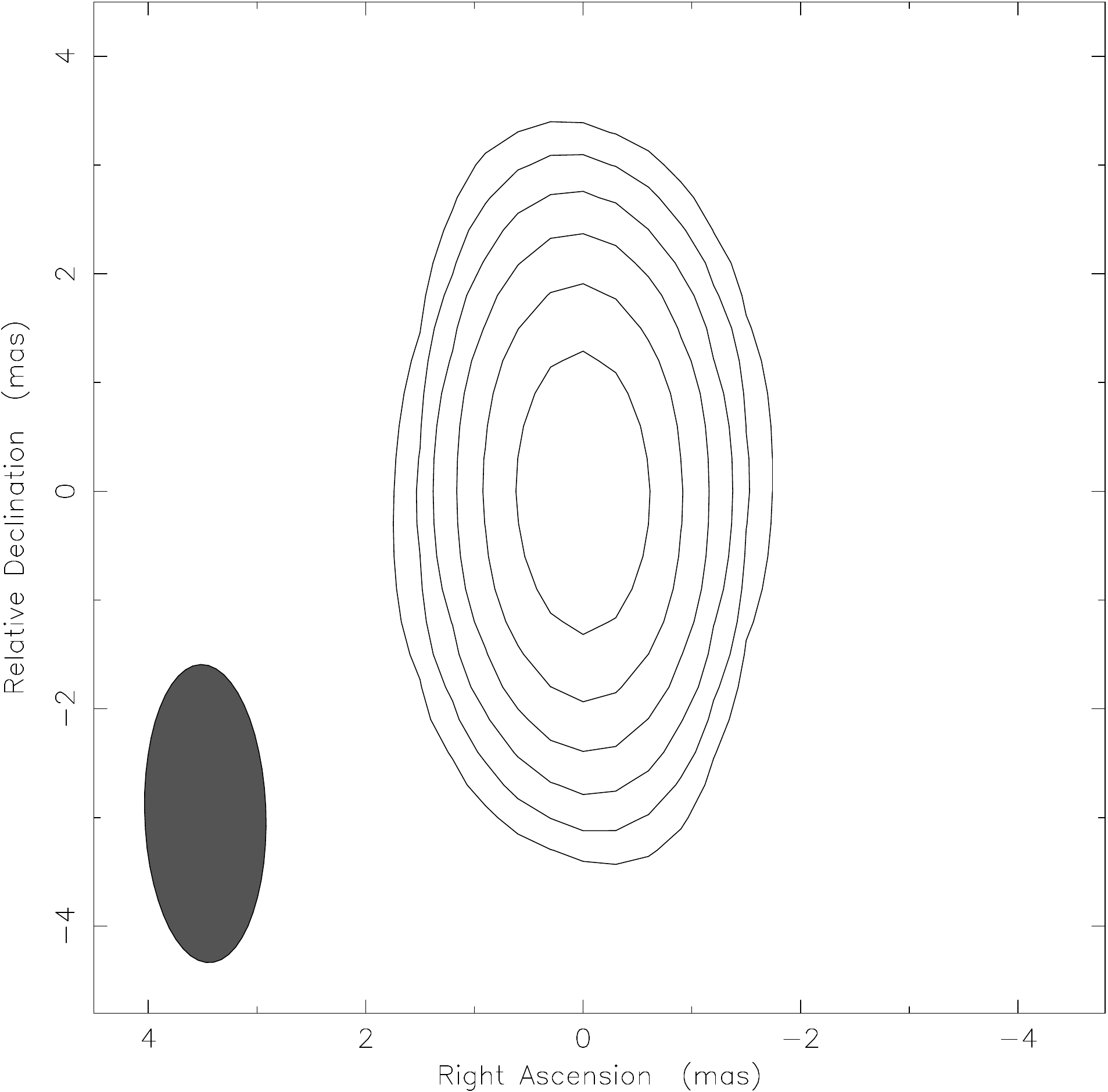}
\label{sgra9-11}}
\qquad
\subfloat[][]{
\includegraphics[width=\textwidth,angle=0, scale=0.22]{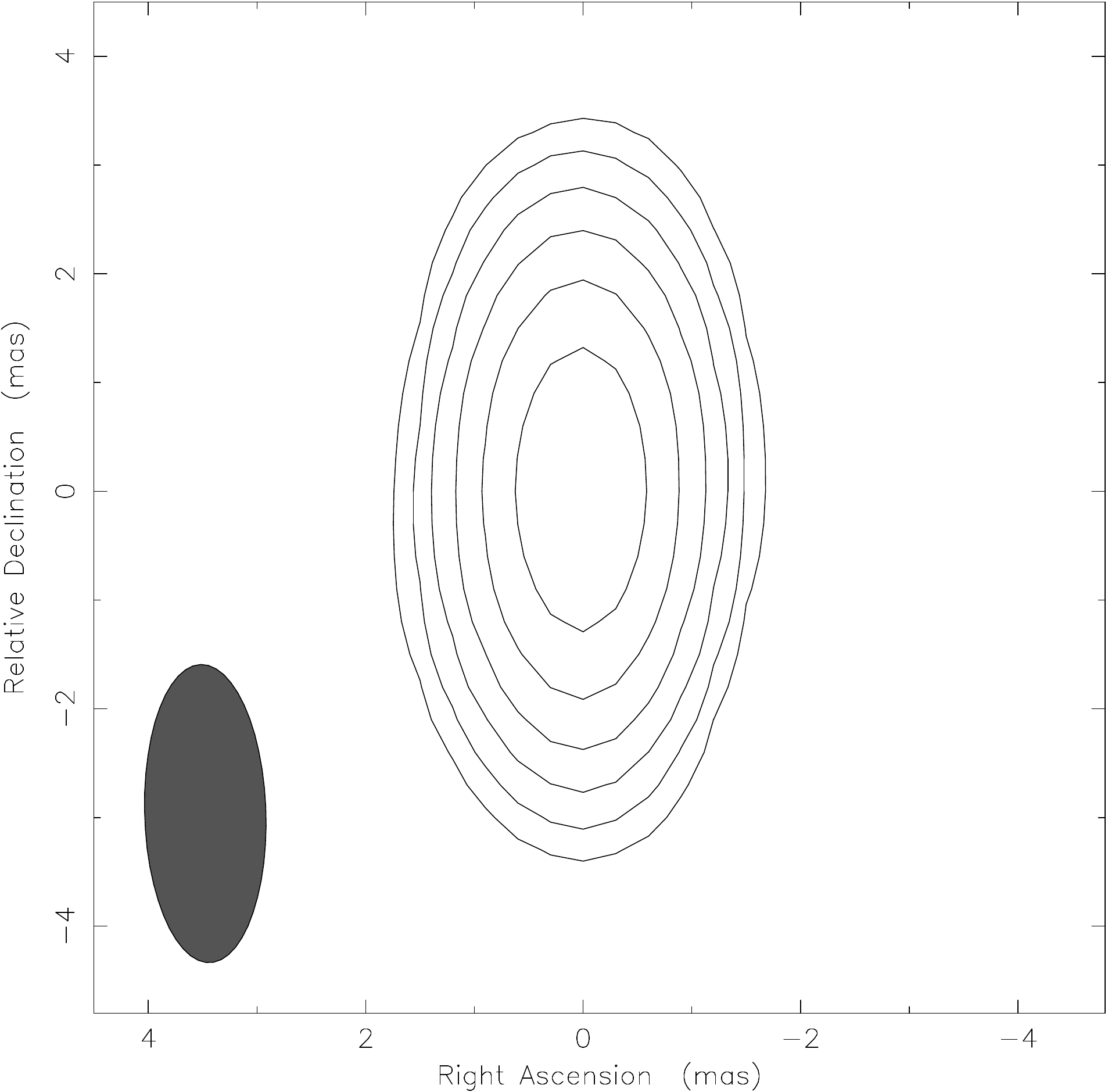}
\label{sgra10-12}}
\caption{\small Two-hour LCP maps of Sgr\,A* observed on May 17 2012. (a) May 17 6-8h (6:04 - 08:00 UT) with a flux of $1.23\pm0.03$. (b) May 17 (7:10 - 09:00 UT) with a  flux of $1.26\pm0.03$. (c) May 17 8-10h (8:14 - 09:58 UT) with a  flux of $1.31\pm0.03$. (d) May 17 9-11h (9:11 - 10:56 UT) with a  flux of $1.41\pm0.03$. (e) May 17 10-12h (10:09 - 12:19 UT) with a  flux of $1.40\pm0.03$. All maps are restored with a beam of 2.74$\times$1.12 at 1.76$^\circ$. The plotted contour levels are 1.73\%, 3.46\%, 6.93\%, 13.9\%, 27.7\%, and 55.4\%.}
\label{sgra2hmaps}
\end{figure*}

\subsection{Calibration and analysis}
\label{calibration}
Calibration and editing of the data was performed using the Astronomical Image Processing System (AIPS). This software package offers standard algorithms for phase and delay calibration and fringe fitting. We also used the \textsc{vlbautil} package, which includes additional calibration scripts. 

The amplitude calibration was performed based on measured antenna system temperatures and gain elevation curves for each station. The atmospheric opacity corrections were determined by plotting the system temperatures against the air mass and fitting the variations (AIPS task: \textsc{APCAL}). Furthermore, corrections for the total electron content (AIPS script: \textsc{vlbatecr}), Earth orientation parameters (AIPS script: \textsc{vlbaeop}), and parallactic angle (AIPS script: \textsc{vlbapang}) were applied. Fringe fitting was made using each target itself as calibrator source and performing a two-point interpolation of delays and delay rates with a solution interval of 30\,s.

Images of Sgr\,A* and the calibrators were produced using the standard hybrid mapping and CLEAN methods of AIPS and DIFMAP (\citealt{shepherd1994}) by averaging over 15\,s at a map resolution of $0.3$\,mas. The beam was enlarged during the iterative amplitude self-calibration process in three steps from one-third over one-half up to its total initial size. Sgr\,A* has only been detected on short baselines between 110-340\,Mega-$\lambda$ (May 16: 220\,Mega-$\lambda$ (FD, KP, LA, OV, PT); May 17: 110\,Mega-$\lambda$ (FD, KP, LA, PT); May 18: 340\,Mega-$\lambda$ (BR, KP, NL, OV, PT)). Marginal detections at longer baselines (MK) were also acquired with fluxes at expected values, but these visibilities could not be modeled accurately and their fluxes increased to implausible values during the self-calibration procedure and were therefore disbanded. The corresponding left-handed circular polarization maps covering a complete observing day of 6\,h are shown in Fig. \ref{be061abc}.

\subsection{Constraints on the data}
\label{constraints}
Observing with the VLBA at 43\,GHz at very low elevations is limited by several effects. The array was designed to operate at longer centimeter wavelengths, steeper gain curves, higher residual pointing, and focus errors. These and other effects are the cause for a deterioration in the data quality (see, e.g., \citealt{lu2011}). Additionally, variable weather conditions and higher atmospheric opacities have a stronger effect at these frequencies than at other radio bands. Even with most cautious observing efforts, Sgr\,A* is susceptible to residual calibration inaccuracies because of its low elevation above the horizon for the VLBA sites located on the northern hemisphere. The low declination of the Galactic center also results in an elliptical UV-coverage and beam size (see Fig. \ref{uvcover_sgra}), which causes lower positional accuracies along the major beam axis.\\
The presented dataset BE061 suffered from many problems of deteriorating
data quality. A variety of technical difficulties during the observations
(NL: not working AC, BR: Wideband upgrade, OV: Saturated system
temperatures) left only up to six stations to be used on each date
for data acquisition. While the only stations present on all
three days are KP, PT, and MK, baselines to the latter unfortunately
did not provide Sgr\,A* detections. Because of changing system
temperature values and other uncorrectable station- and scan-based
flux errors, the calibrators 3C\,345 and 3C\,279 had to be excluded
from most of the analysis. The remaining calibrator NRAO\,530
provided acceptable data quality and shows a constant flux of
(2.42$\pm$0.04)\,Jy over the whole experiment.\\
On May 18, the flux values of Sgr\,A* dropped by $\approx$50\% on
all baselines after 10:00\,UT. These visibilities are therefore
considered to be faulty and were excluded from this discussion.

\begin{figure*}[]
\centering
\subfloat[][]{
\includegraphics[width=\textwidth,angle=0, scale=0.22]{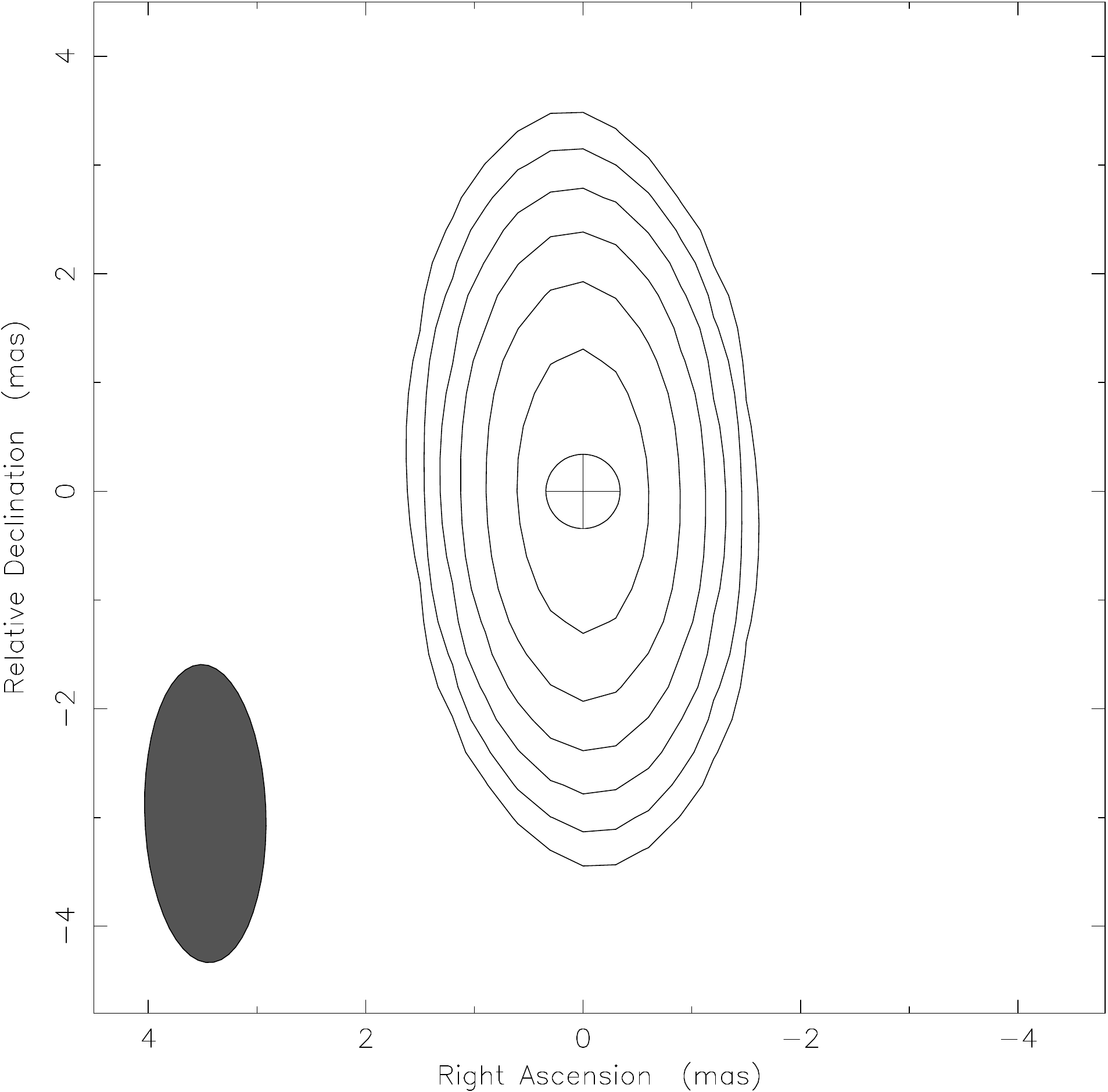}
\label{sgramodelfit6-8}}
\qquad
\subfloat[][]{
\includegraphics[width=\textwidth,angle=0, scale=0.22]{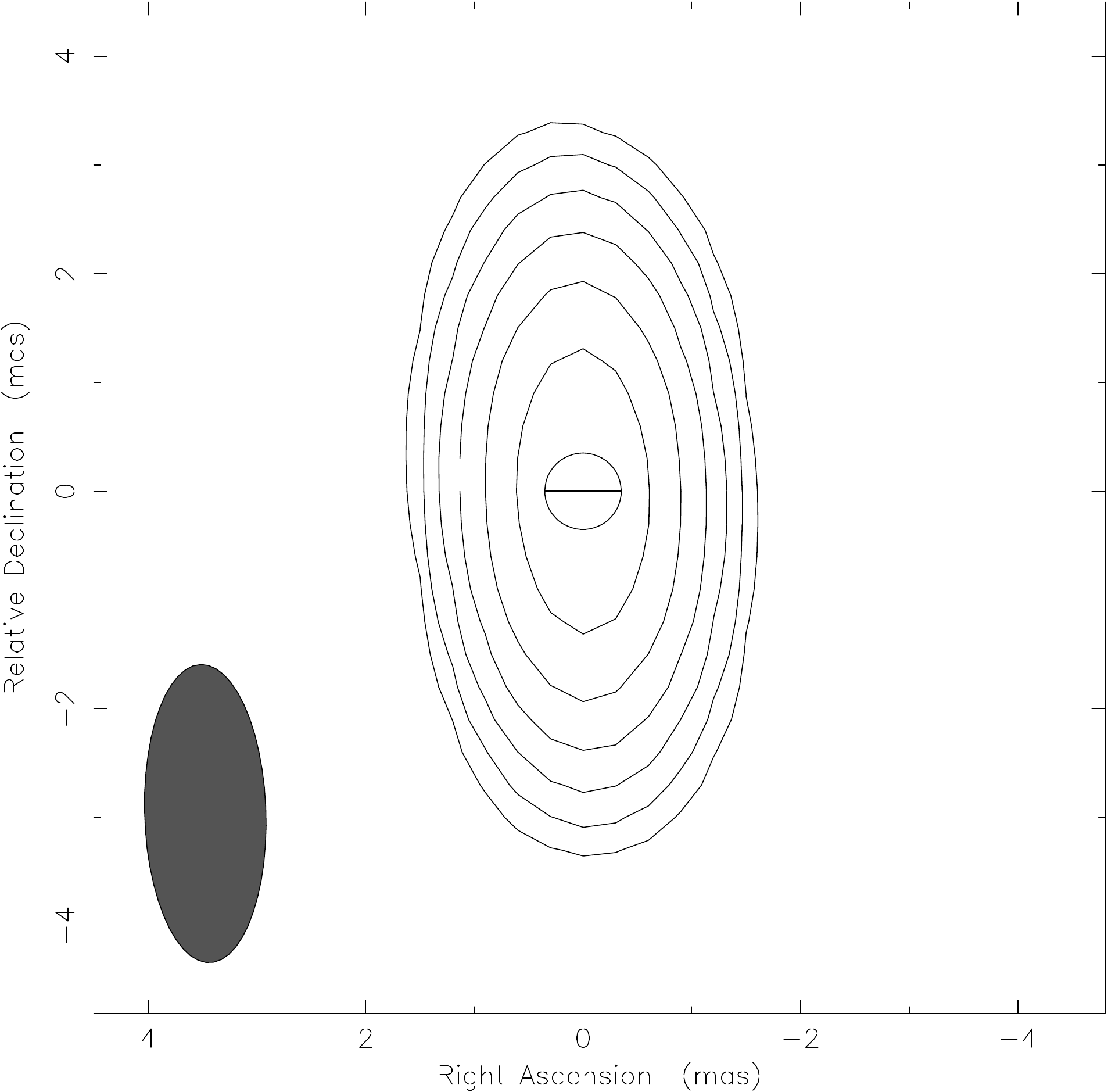}
\label{sgramodelfit7-9}}
\qquad
\subfloat[][]{
\includegraphics[width=\textwidth,angle=0, scale=0.22]{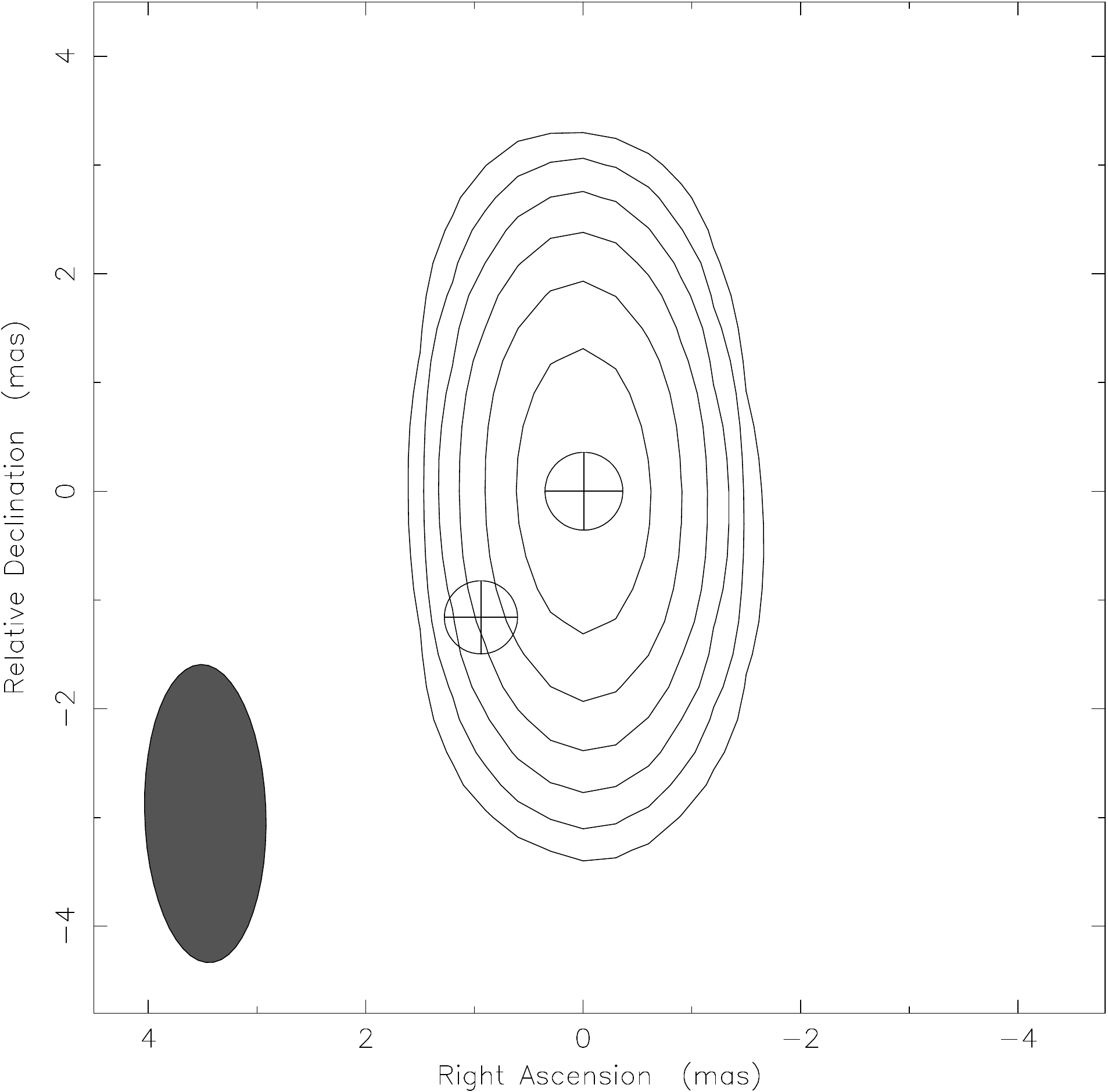}
\label{sgramodelfit8-10}}
\qquad
\subfloat[][]{
\includegraphics[width=\textwidth,angle=0, scale=0.22]{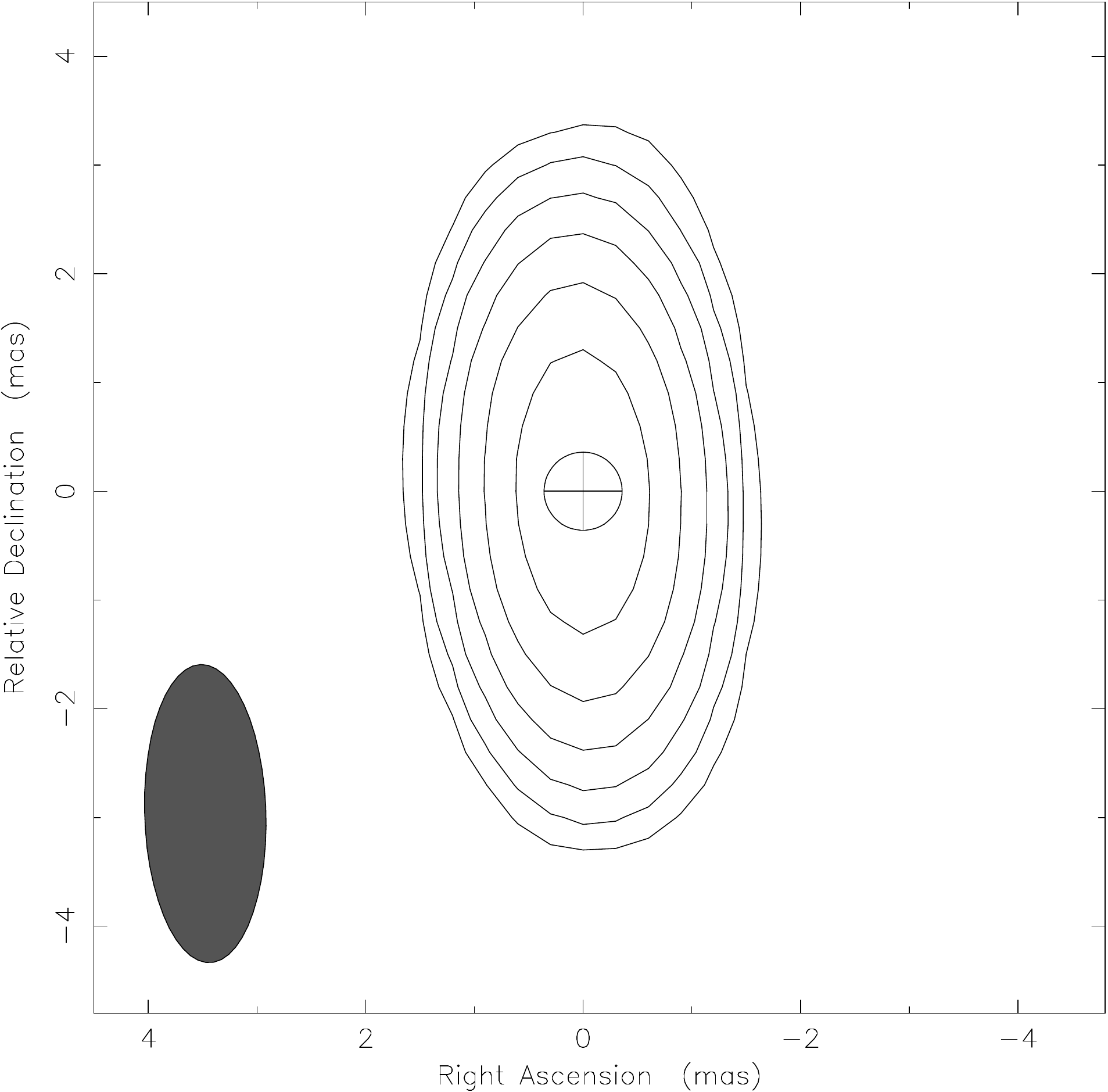}
\label{sgramodelfit9-11}}
\qquad
\subfloat[][]{
\includegraphics[width=\textwidth,angle=0, scale=0.22]{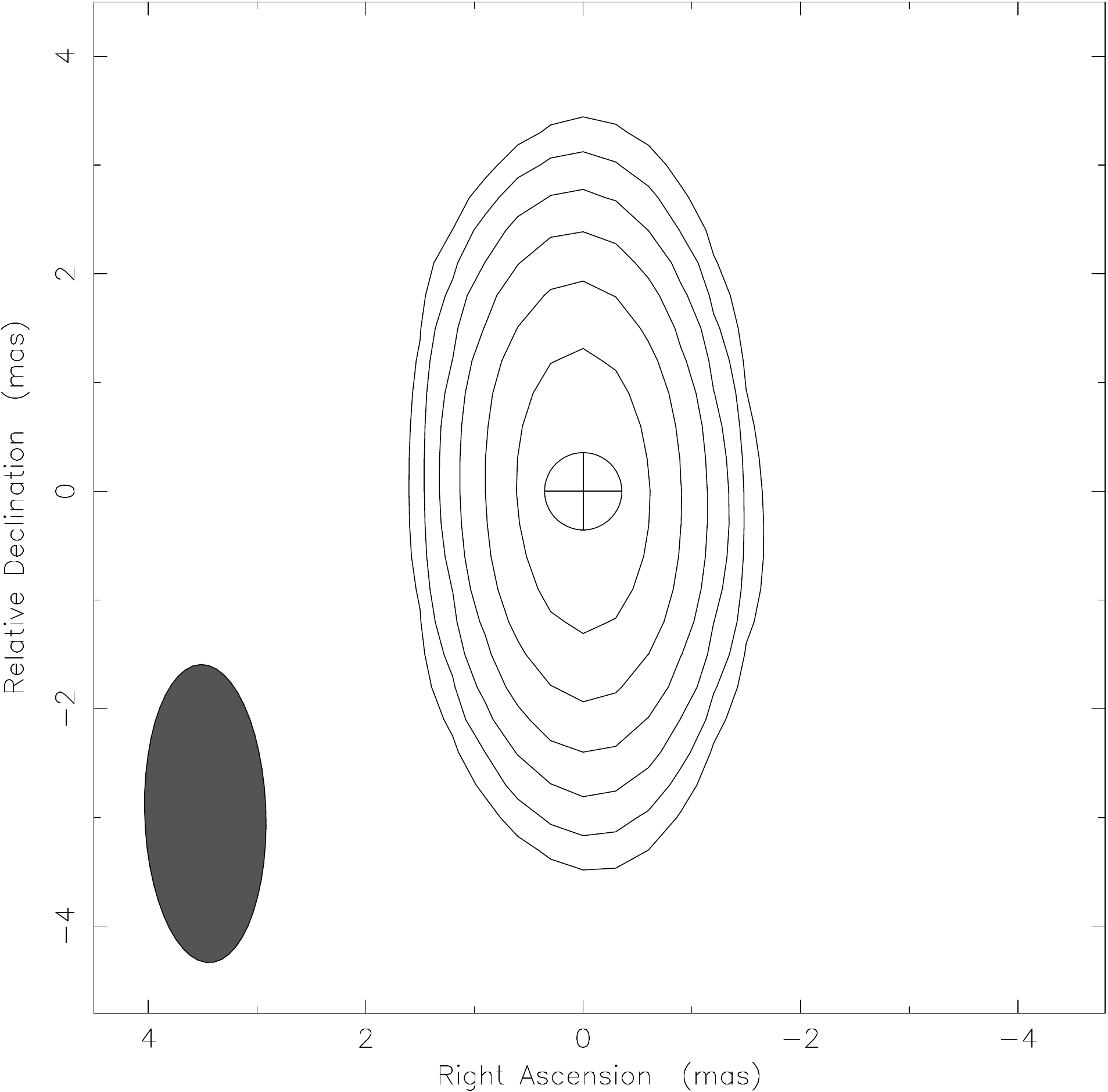}
\label{sgramodelfit10-12}}
\caption{\small 2 hour model fits of Sgr\,A* on May 17 2012. Extended morphology during flare is best fitted by a two component model. (a) May 17 6-8h (6:04 - 08:00 UT), flux: $1.21\pm0.03$. (b) May 17 (7:10 - 09:00 UT), flux: $1.21\pm0.03$. (c) May 17 8-10h (8:14 - 09:58 UT), flux: $1.26\pm0.03$. (d) May 17 9-11h (9:11 - 10:56 UT), flux: $1.29\pm0.03$. (e) May 17 10-12h (10:09 - 12:19 UT), flux: $1.28\pm0.03$. All maps are restored with a beam of 2.74$\times$1.12 at 1.76$^\circ$. The plotted contour levels are 1.73\%, 3.46\%, 6.93\%, 13.9\%, 27.7\% and 55.4\%.}
\label{sgramodelfit}
\end{figure*}

\section{Results}
\label{results}

\subsection{Flux density}
\label{fluxdensity}
 
On May 17, 2012 the VLT detected a NIR flare peaking at 5:30\,UT (Fig. \ref{NIR_lightcurve}). We have plotted the discrete Fourier transform (DFT) of the complex visibilities for Sgr\,A* and the calibrator NRAO\,530 provided by the AIPS task \textsc{DFTPL} (Fig. \ref{be061_sgra_nrao530_7mm_lightcurve}). On this date an increase of its flux density from 0.97\,Jy to 1.17\,Jy at $\sim$9:30\,UT has been observed (see Fig. \ref{be061b_sgra_detrend}). This intensity was derived from uncalibrated data and is free of any errors that might arise during calibration. This therefore
is the most unbiased way to determine relative flux trends that
are introduced by observational effects. The light curve of Sgr\,A* was corrected for the intensity fluctuations of NRAO\,530 by determining the difference of every data point from the mean value in percent and subtracting these as percentages of its mean value from Sgr\,A*. This produces the detrended light curve of Sgr\,A* presented in Fig. \ref{be061b_sgra_detrend}. Based on the flux measurements of NRAO\,530, we adopted an error of 1.7\% for all amplitudes within this dataset. It would be better to use the flux densities of closer calibrators such as the phase reference sources, but these are too weak to be used for an accurate flux estimation and are therefore not regarded in this work.

For further analysis the presented 7\,mm data on this date were split into five overlapping 2\,h parts and then mapped individually (see Fig. \ref{sgra2hmaps}). Because of technical problems during the observations, only LCP data provided acceptable data quality and therefore RCP had to be disbanded. We observe increasing flux densities at 7\,mm peaking between 9:00-11:00 UT (see Fig. \ref{7mm_lightcurve}). The flux density on May 16 and May 18 remained constant at values of (1.14$\pm$0.02)\,Jy and (1.0$\pm$0.2)\,Jy, respectively. May 18 was only considered until 10:00\,h UT because
of observational problems.  While Sgr\,A* shows constant flux densities during its quiescent states on May 16 and 18, its intensity increases from $1.23$\,Jy (6:00-8:00 UT) to $1.41$\,Jy (9:00-11:00 UT) on May 17, which agrees well with the fluxes observed at the shortest baselines before self-calibration and is well above the quiescent flux values reported on May 16 and 18. This intensity variability is seen on independent baselines, which excludes the possibility that station-based errors affect the measured flux densities. Each map was produced using the same beam and map size to achieve consistent and comparable images. The calibrated fluxes of NRAO\,530 show systematic station-based correction factors of $\sim$18$\%$ compared to its uncalibrated values, which agrees well with the values found for Sgr\,A*. We  therefore considered them to be corrected by the presented calibration. For this analysis, we excluded some poor visibilities from PT at the start and end of the experiment. The FD station inherits an uncorrectable offset with respect to the other stations and was therefore completely excluded.

Additionally, Fig. \ref{sgra2hmaps} shows Sgr\,A* as a point-like source during the initial quiescent state and then developing an extended feature of $\sim$(1.7$\pm$0.3)\,mas toward the southeast during its flaring state. The apex of this change of morphology appears shortly (8:00-10:00\,h UT) before the peak of the flux density at 9:00-11:00\,h UT. To test the validity of this feature and rule out clean-window-biased and faulty detected components, we performed different cleaning methods by changing the beam sizes and areas covered by clean windows on all maps during the hybrid mapping process, which all developed a somewhat pronounced extended feature at the same position. Model fitting by placing circular Gaussian components to the UV-data at the regions of expected flux, as suggested by the clean maps, and solving for size, radius, and positional degree yields a reliable test of the observed source morphology. Figure \ref{sgramodelfit} shows the resulting best-fit model components. The presented maps were modeled using components well above the scattering size at 7\,mm and favor a two-component model during the periods of higher flux densities, as intended by the clean maps. The $\chi^2$-value of this model fit represents the goodness of the model fitting
of the UV-data and is therefore a measure for the quality of the input model. Trying to model a single component for the time of the most asymmetric maps (8:00-10:00h UT) produces higher reduced $\chi^2$ values than a two-component model (two components: $\chi^2=0.34$, single component: $\chi^2=0.54$).

Analyzing the corresponding right-handed circular polarization map reveals a similar feature. The positions of the detected delta components in the left- and right-handed circular polarization maps correspond to a position of the secondary component at a radial separation of $(1.7\pm0.3)$\,mas (LCP) and $(1.8\pm0.2)$\,mas (RCP) at position angles of $(126\pm16)^{\circ}$ (LCP) and $(138\pm9)^{\circ}$ (RCP). These values are consistent within their error limits and are another indication that the secondary component that
is present at this position is a real feature and not a data artifact.

\subsection{Closure phases}
\label{closurephases}
\begin{figure}[!htb]
\centering
\subfloat[][]{
\includegraphics[width=\textwidth,angle=0, scale=0.35]{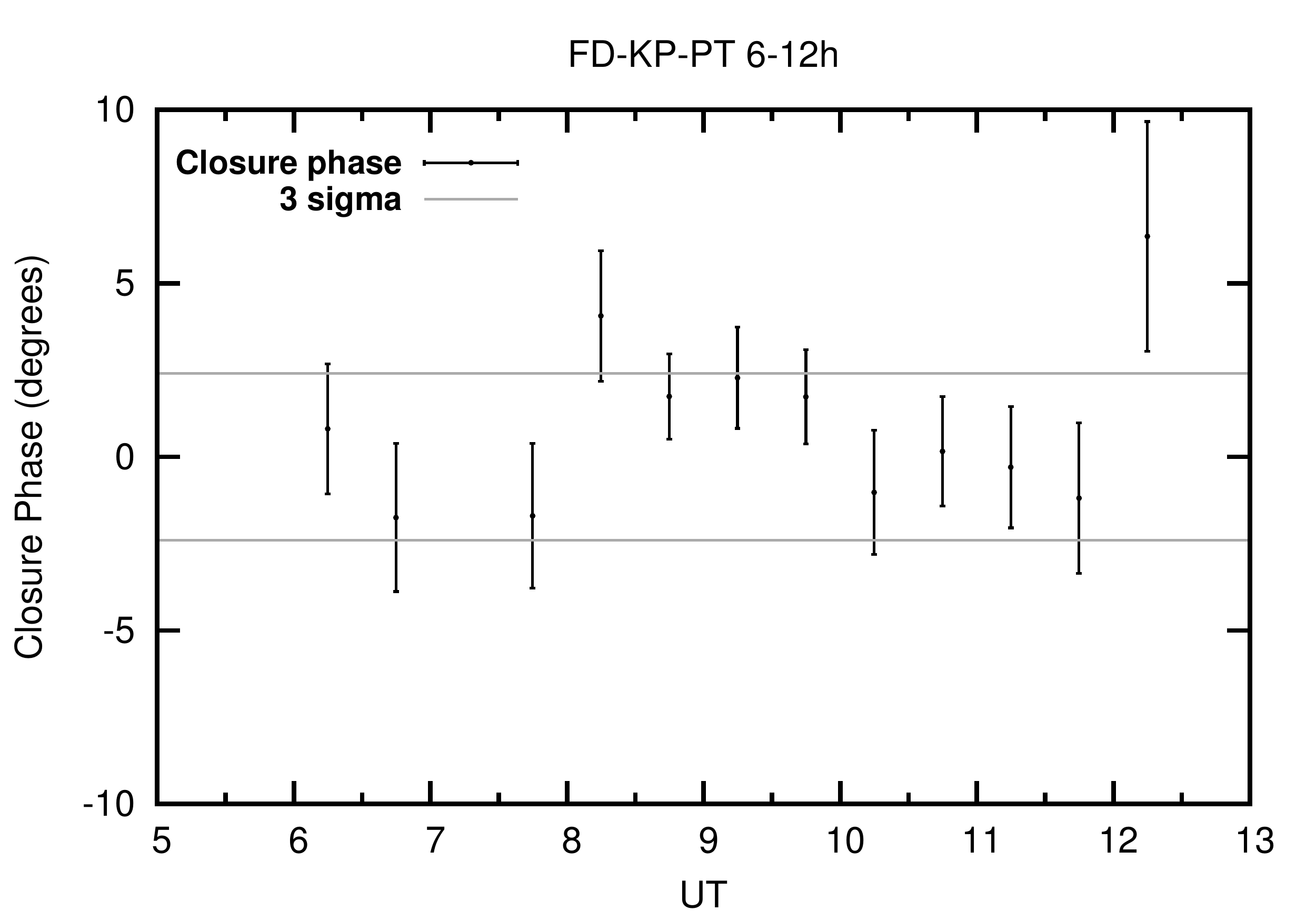}
\label{FD-KP-PT_cp_6-12}}
\qquad
\subfloat[][]{
\includegraphics[width=\textwidth,angle=0, scale=0.35]{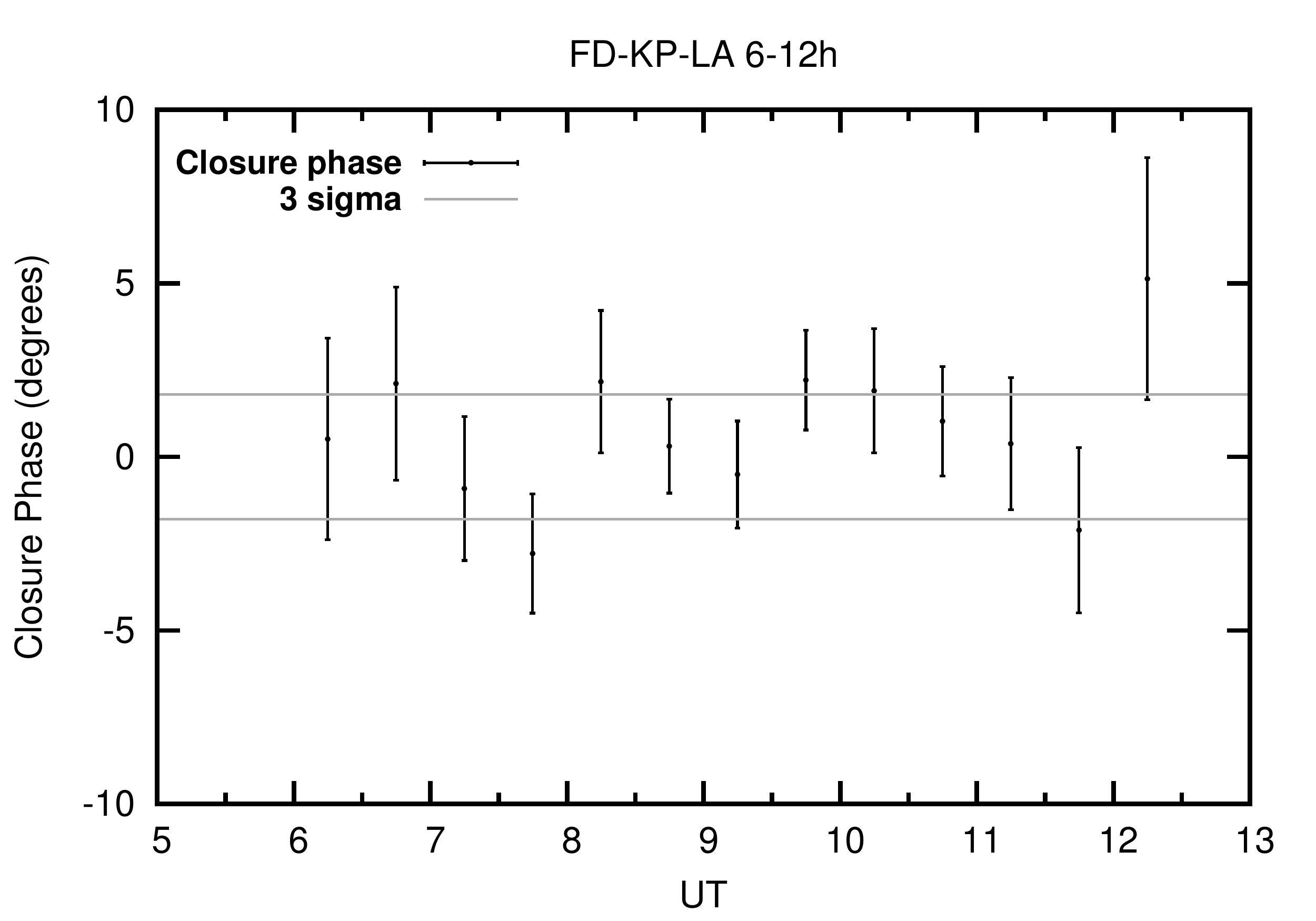}
\label{FD-KP-LA_cp_6-12}}
\qquad
\subfloat[][]{
\includegraphics[width=\textwidth,angle=0, scale=0.35]{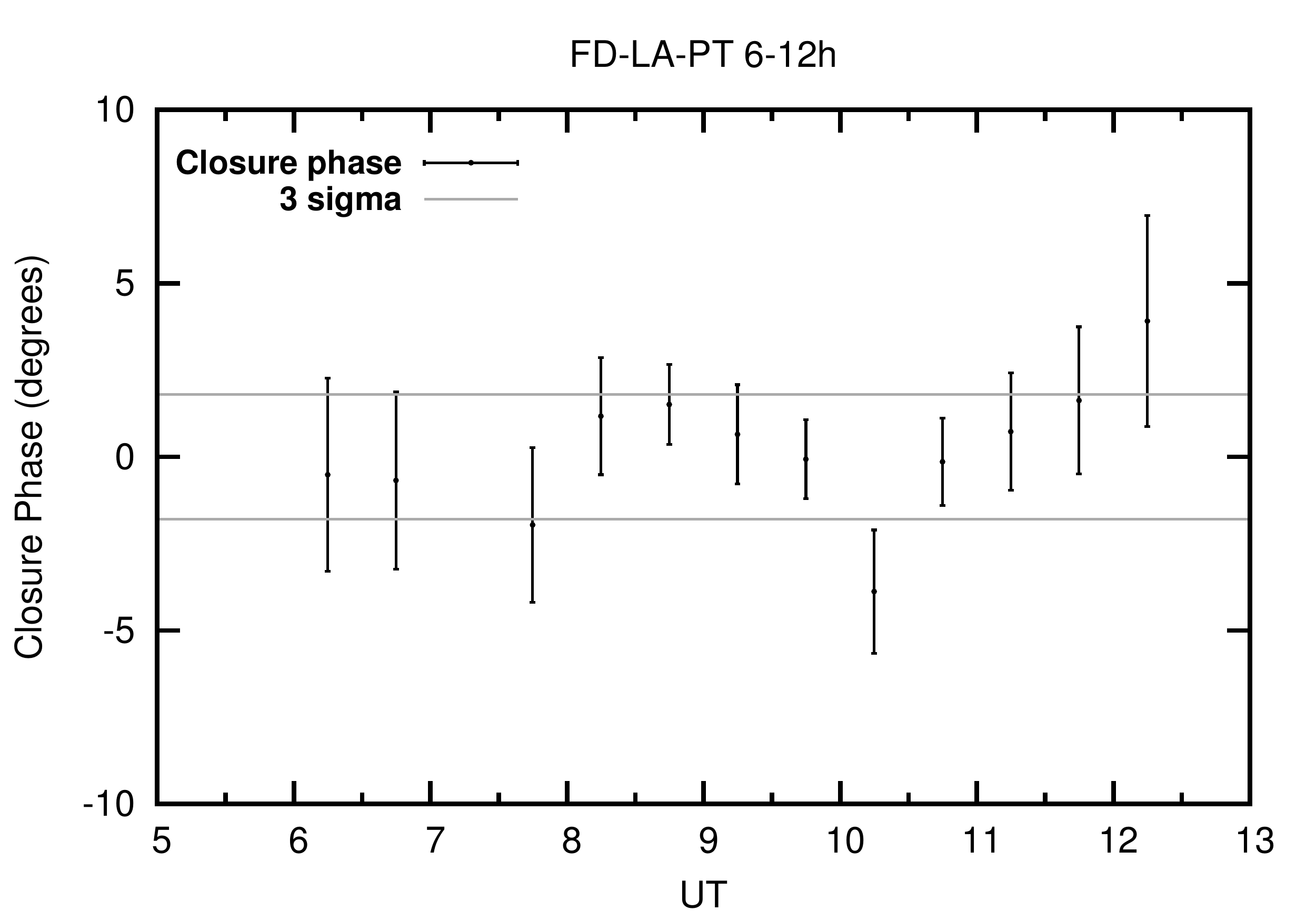}
\label{FD-LA-PT_cp_6-12}}
\qquad
\subfloat[][]{
\includegraphics[width=\textwidth,angle=0, scale=0.35]{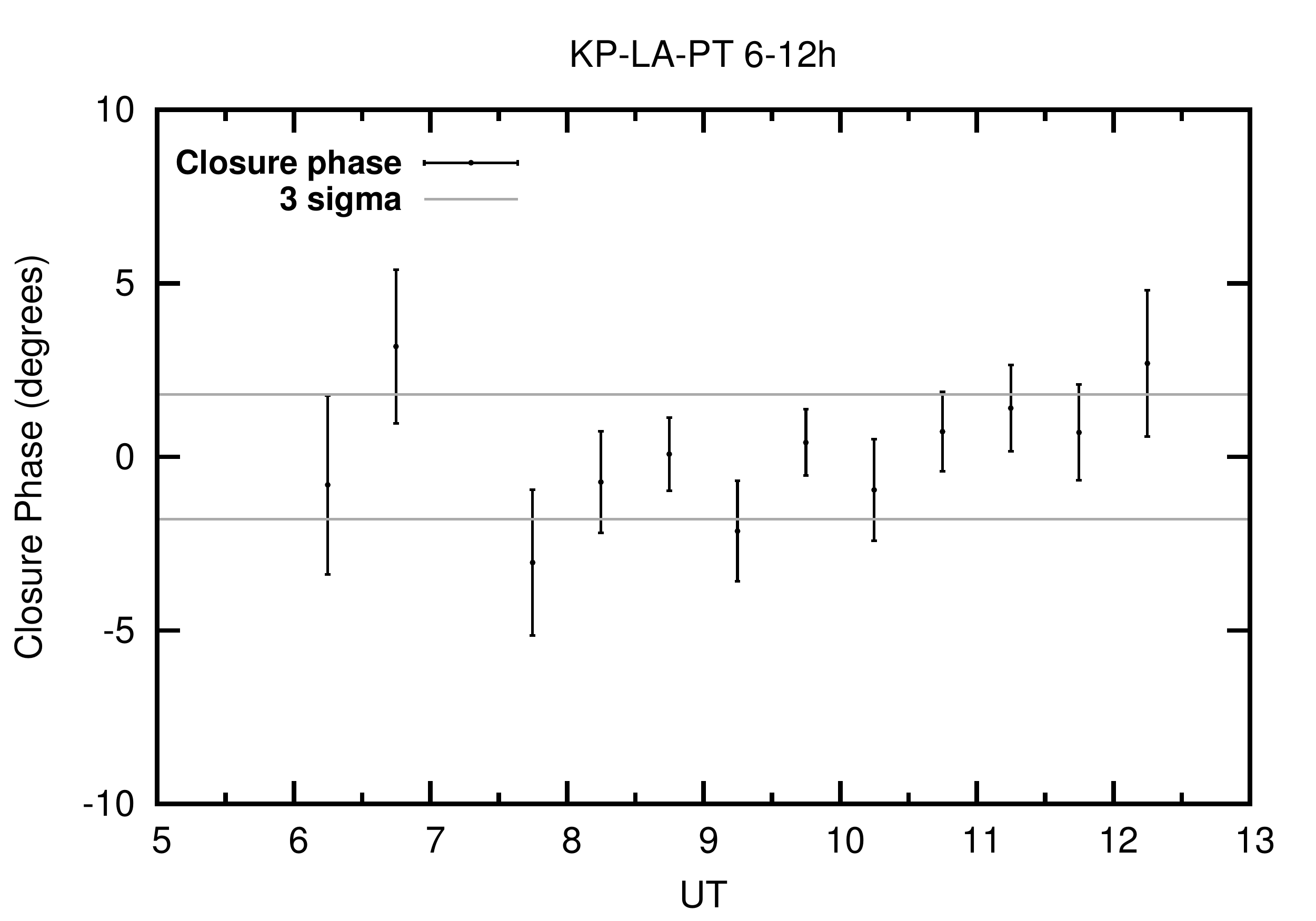}
\label{KP-LA-PT_cp_6-12}}
\caption{Closure phases of all closure triangles on May 17 6:00-12:00\,h UT with plotted standard deviation levels. (a) FD-KP-PT, (b) FD-KP-LA, (c) FD-LA-PT, (d) KP-LA-PT. We show standard errors provided by standard deviation.}
\label{be061b_cp}
\end{figure}

A good quantity on which to test the symmetry of a compact radio source are the closure phases, which are the sum of three baseline phases in a triangle of antennas. These phases are independent of all station-based phase errors because they are a phase quantity of the complex visibilities. A point-symmetric source would have closure phases of zero, while any deviation from its symmetry would cause non-zero values.
In Fig. \ref{be061b_cp} we plot closure phases for May 17 extracted from 30-min averaged UV-data. A mean value was determined for each set of closure triangles in dual polarizations as well as an average mean phase for all triangles of $(0.5\pm0.2)^{\circ}$ (LCP) and $(0.0\pm0.1)^{\circ}$ (RCP) (see Table \ref{be061b_cp_tab}). The values for all triangles and polarizations are all equal within their error limits. Because the station arrays changed on May 18, there are no high-quality closure phases that can be compared to the previous days, and they can therefore not be used in this analysis.

Of most interest is the closure triangle FD-KP-PT, which shows non-zero closure phases at 8:00-10:00\,h UT on May 17. While no triangle offers values exceeding $3\,\sigma$, there are several values higher than $2\,\sigma$. These excessive values coincide with the observed change in morphology at this time (see Fig. \ref{sgra8-10}). This trend is not reproduced by the other triangles, which also show some closure phases above the two-sigma range, but in a more random fashion.

To further investigate this effect, we tried to simulate the observed closure phase structure using the Caltech VLBI analysis program \textsc{FAKE} (\citealt{pearson1991}). \textsc{FAKE} is limited to single polarization and a single IF, but since the presented maps of this work are LCP measurements, and there is no a priori reason for simulated RCP values to be different,
and this should not change the result. We simulated several datasets using different input models (see Table \ref{be061b_fake_cp_tab}) of the source and parameters similar to the presented measurement campaign. The summary of all simulated closure phase values can be found in Table \ref{be061b_fake_cp_tab}. 

Starting with two components separated by 1.5\,mas at $140^\circ$ east of north (central component: 1.55\,Jy, secondary component: 0.02\,Jy), as suggested by the presented modelfit, we tried to reproduce the observed closure phase values by changing the position of the secondary component to 0.7\,mas, 0.3\,mas, and finally a single central component. All of these simulations for the time range of maximum source asymmetry (8-10\,h on May 17) produced higher closure phases than observed. Therefore, we changed the method of error application to the data. The applied errors for the simulations were first determined from Tsys, station diameters, station efficiencies, and pointing efficiency, and second by changing the amplitude of an applied additive Gaussian noise (\textsc{erradd}). \textsc{erradd} specifies the amplitude of additive Gaussian noise to a set of stations. The error of stations I and J is computed by $(\textsc{erradd(I)} \times \textsc{erradd(J)})^\frac{1}{2}$. The simulated datasets also show higher errors in the range of 0.8 to 6.7 and are all zero within these limits, while a perfect simulation without any noise produces very low errors and values similar to those that are observed. To understand this behavior, additional datasets were simulated using different seeds, which were used to generate random numbers for error application to the phases and amplitudes in \textsc{FAKE}. These simulations show that depending on the chosen seed, the mean closure phases for the presented triangles can differ by $\sim$10$^\circ$ for each closure triangle, and since the observed effect on closure phases is about $\sim$5$^\circ$ , it is not possible to accurately simulate the phases of the presented data set. 

In this context, it can be expected that a secondary component at $140^\circ$ will produce higher closure phases on closure triangles with an axis of maximum resolution close to this angle. Therefore, FD-KP-LA, FD-KP-PT, and FD-LA-PT should show the highest closure phase values since their maximum resolution angles are at $97^\circ$, $101^\circ$, and $125^\circ$ and thus closer to the asymmetric feature than KP-LA-PT ($65^\circ$). This is well reproduced by the perfect simulation without errors, but it can change significantly depending on the seed value for randomization. Even though for all simulations the highest deviation from zero is present in one of the three favored triangles, a chosen seed was capable of producing a mean value of the closure phase for KP-LA-PT of $7.2^\circ$ , which is the second highest value among triangles within this simulation. The observed closure phases of our dataset show the highest closure angle for FD-KP-PT of $1.9^\circ$ , but it also has its second highest value for KP-LA-PT. Even though these values might be similar to the expected trend, as suggested by the angle of maximum resolution and the perfect simulation, with the given error limits, we can not place reliable constraints on this hypothesis.

To test the probability of detecting the observed source structure with the given array, we iterated several simulations with different Gaussian noise values. Starting at similar values to our observations and increasing the noise level in 10\% steps revealed that the two-component source structure can be reproduced up to Gaussian noise values of \textsc{erradd}=0.16 with a $\chi^{2}$=1.22. To test the possible effect of noise being the cause of the secondary component in Sgr\,A*, we performed a similar simulation by generating a single-component dataset and increasing the Gaussian noise to values of up to \textsc{erradd}=0.80 and checked the resulting visibilities for indications of a possibly appearing secondary component generated by random noise. The resulting visibilities showed no evidence of a secondary component for all simulated values.

\begin{figure}[!htb]
\centering
\includegraphics[width=\textwidth,angle=-90, scale=0.35]{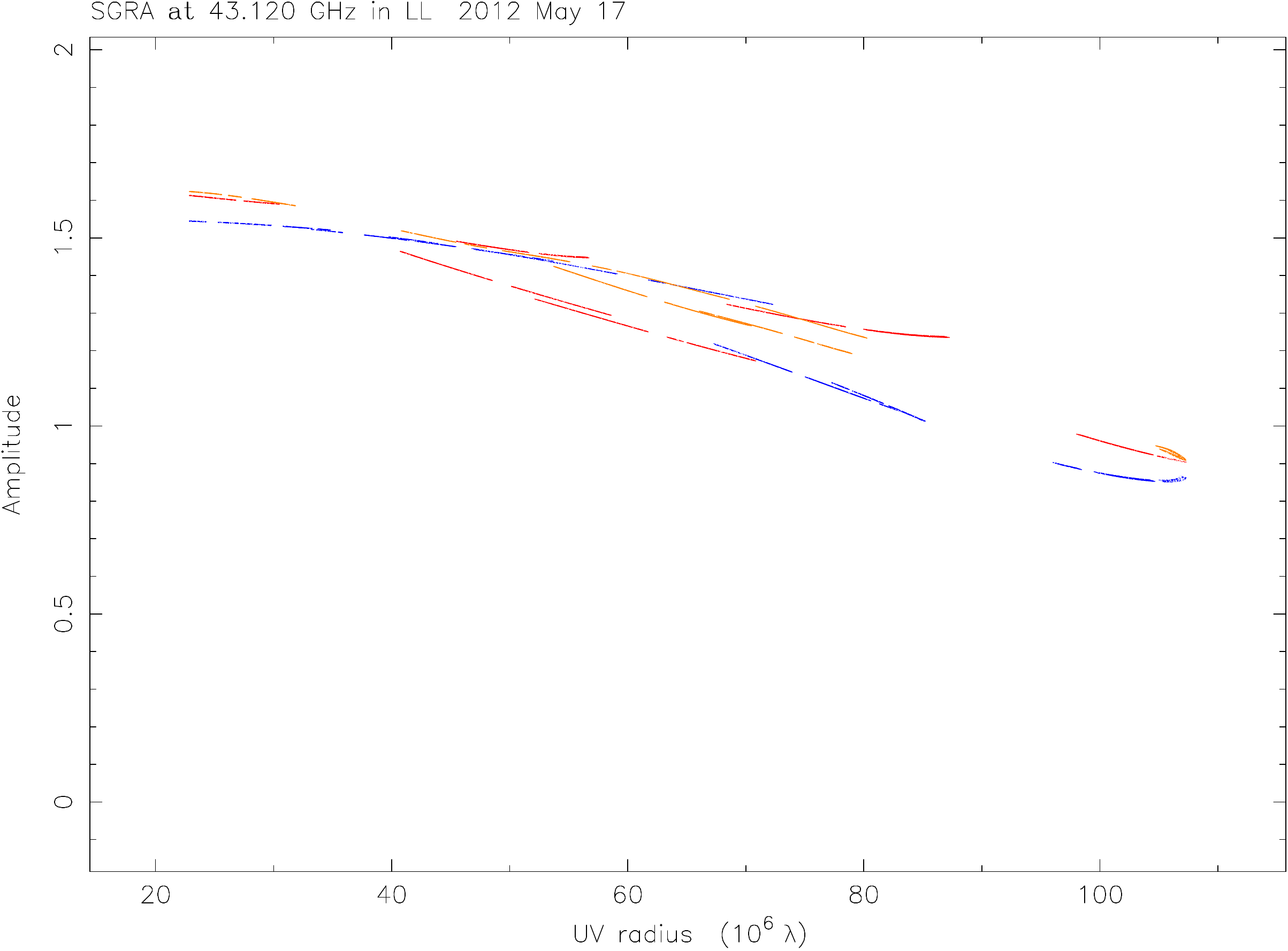}
\caption{\small Radial plot of the models obtained during the time of highest flaring activity of Sgr\,A* on May 17, 2012. The colors correspond to the maps of 8:00-10:00h UT (red), 9:00-11:00h UT (orange), and 10:00-12:00h UT (blue).}
\label{be061b_sgra_radpl}
\end{figure}

By examining the time range of highest flare activity ranging from 8:00-12:00\,h UT on May 17, 2012, we obtained a radial plot of the model generated for the clean maps of Sgr\,A* on May 17 (see Fig. \ref{be061b_sgra_radpl}). It is noticeable that the most point-like map on 10-12\,h (blue) does not inherit a strong secondary component in its models for UV-ranges between 40 - 80\,M$\lambda$, while the two models for the more asymmetric maps during 9-11\,h (orange) and especially during 8-10\,h (red) clearly show secondary model features. This means that the data were best fit by two components for this UV-range, and it is another indication for the existence of an off-core feature.
All presented independent tests fit consistently with the picture that Sgr\,A* experienced a change in morphology on May 17 8:00-10:00\,h UT.
\begin{figure}[!h]
\centering
\includegraphics[width=\textwidth,angle=0, scale=0.3]{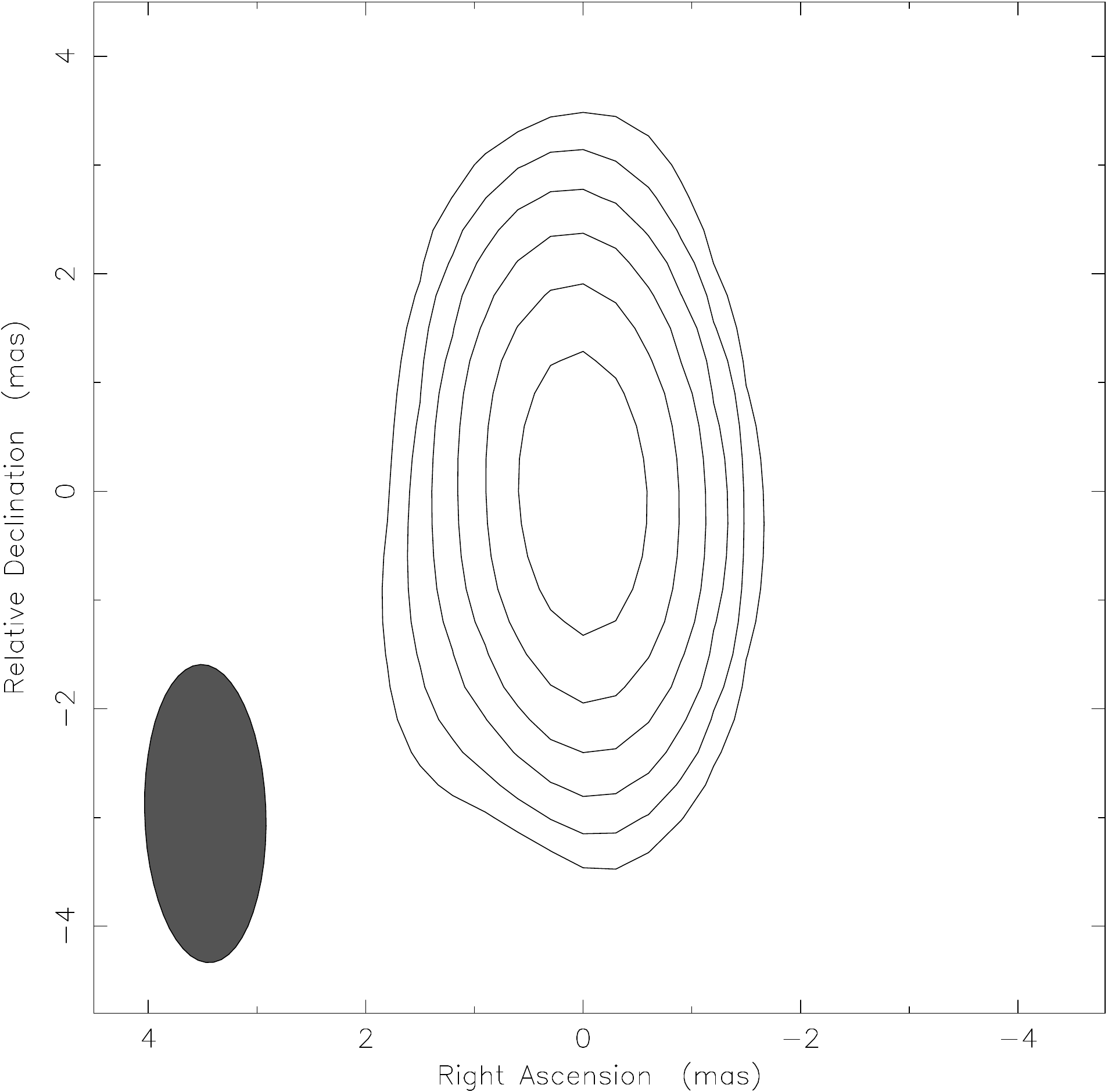}
\caption{\small RCP map of Sgr\,A* on May 17, 2012 (8:00-10:00h UT). The map was convolved with a beam of 2.74$\times$1.12 at 1.76$^\circ$. Contour levels are 1.73\%, 3.46\%, 6.93\%, 13.9\%, 27.7\%, and 55.4\% of the peak flux density of 1.5\,Jy/beam.}
\label{be061b_sgra_RR_8-10h}
\end{figure}

\section{Discussion}
\label{discussion}
Several sources claim quasi-periodic NIR flares appearing on timescales ranging from 1-2 hours (main flares) down to 7-10 minutes (sub-flares) (\citealt{genzel2003,aschenbach2004,eckart2006c}). Since the discovery of these flares, many models have been discussed. These models can be distinguished by a wide range of properties such as their periodicity or the source morphology (see Sect. \ref{introduction}).\\

\begin{table*}[]
\centering
\resizebox{0.78\textwidth}{!}
{\begin{minipage}{\textwidth} 
\begin{tabular}{lcccccc}
\hline
\hline Time & & FD-KP-LA & FD-KP-PT & FD-LA-PT & KP-LA-PT & Sample mean \\ 
\hline 8-12 (May 16, 2012) & LCP: & $(0.2\pm0.5)^{\circ}$ & $(-0.3\pm0.6)^{\circ}$ & $(-0.6\pm0.4)^{\circ}$ & $(0.0\pm0.4)^{\circ}$ & $(0.2\pm0.3)^{\circ}$\\
 & RCP: & $(-2.0\pm1.5)^{\circ}$ & $(-3.4\pm1.1)^{\circ}$ & $(-1.2\pm0.8)^{\circ}$ & $(0.2\pm0.6)^{\circ}$ & $(-1.6\pm0.8)^{\circ}$\\
 \\
 6-12 (May 17, 2012) & LCP: & $(0.7\pm0.6)^{\circ}$ & $(0.9\pm0.8)^{\circ}$ & $(0.2\pm0.6)^{\circ}$ & $(0.1\pm0.6)^{\circ}$ & $(0.5\pm0.2)^{\circ}$ \\
  & RCP: & $(0.1\pm0.7)^{\circ}$ & $(0.1\pm0.7)^{\circ}$ & $(-0.1\pm0.5)^{\circ}$ & $(-0.2\pm0.3)^{\circ}$ & $(0.0\pm0.1)^{\circ}$\\
\hline Two-hour map \\ 
\hline 8-10 (May 17. 2012) & LCP: & $(0.6\pm1.2)^{\circ}$ & $(1.9\pm1.1)^{\circ}$ & $(0.2\pm0.8)^{\circ}$ & $(-1.1\pm0.8)^{\circ}$ & $(0.4\pm0.7)^{\circ}$ \\
 & RCP: & $(-0.8\pm0.9)^{\circ}$ & $(-1.1\pm0.3)^{\circ}$ & $(0.0\pm0.3)^{\circ}$ & $(0.3\pm0.8)^{\circ}$ & $(0.4\pm0.4)^{\circ}$\\
\hline 
\end{tabular}
\centering
\caption{\small Summary of closure phases for all closure triangles on May 16 and 17.} 
\label{be061b_cp_tab}
\end{minipage}}
\end{table*}

\begin{table*}[]
\centering
\resizebox{0.8\textwidth}{!}
{\begin{minipage}{\textwidth} 
\begin{tabular}{lccccc}
\hline
\hline Models & FD-KP-LA & FD-KP-PT & FD-LA-PT & KP-LA-PT & Sample mean \\ 
\hline Radius = 1.5\,mas & 2.3$\pm$5.5 & -2.0$\pm$2.7 & 2.7$\pm$4.6 & -1.6$\pm$2.0 & 0.4$\pm$1.3 \\ 
Radius = 0.7\,mas & 1.2$\pm$5.5 & -2.8$\pm$2.7 & 2.4$\pm$4.7 & -1.6$\pm$2.0 & -0.2$\pm$1.3 \\
Radius = 0.3\,mas & 1.0$\pm$5.5 & -2.9$\pm$2.7 & 2.3$\pm$4.6 & -1.6$\pm$2.0 & -0.3$\pm$1.2 \\
Single component & 2.7$\pm$2.0 & -2.5$\pm$1.6 & 2.5$\pm$0.8 & -2.7$\pm$1.7 & 0$\pm$1.6 \\
\hline Seeds\\
\hline Seed = 3456757 & 2.4$\pm$5.5 & -2.0$\pm$2.7 & 2.8$\pm$4.6 & -1.6$\pm$2.0 & 0.4$\pm$1.3 \\
Seed = 3000000 & -1.2$\pm$4.6 & 1.8$\pm$1.6 & -3.9$\pm$3.3 & -0.9$\pm$5.4 & -1.1$\pm$1.2 \\
Seed = 1278562 & 8.6$\pm$1.9 & 1.9$\pm$2.4 & -0.4$\pm$3.8 & -7.2$\pm$6.7 & 0.7$\pm$3.3 \\
\hline Errors\\
\hline Perfect & 1.35$\pm$0.09 & 1.00$\pm$0.06 & -0.44$\pm$0.04 & 0.08$\pm$0.01 & 0.7$\pm$0.3 \\
Erradd = 0.02 & 2.4$\pm$5.6 & -1.6$\pm$2.5 & 2.5$\pm$4.6 & -1.5$\pm$1.8 & 0.5$\pm$1.2 \\
Erradd = 0.01 & 2.5$\pm$5.7 & -1.5$\pm$2.5 & 2.5$\pm$4.6 & -1.5$\pm$1.7 & 0.5$\pm$1.2 \\
Erradd = 0.005 & 2.5$\pm$5.7 & -1.5$\pm$2.5 & 2.4$\pm$4.6 & -1.5$\pm$1.7 & 0.5$\pm$1.2 \\
\hline
\end{tabular}
\centering
\caption{\small Summary of closure phases for all simulated closure triangles on May 17 8:00-10:00\,h UT.} 
\label{be061b_fake_cp_tab}
\end{minipage}}
\end{table*}

Sgr\,A* shows variable flux densities from the radio to NIR and X-ray wavelengths. First evidence of correlated multi-frequency flare activities has been made in the X-ray and radio bands (\citealt{zhao2004}). Related flares at NIR and sub-mm frequencies have been detected by \cite{eckart2008c} and \cite{yusef-zadeh2009}. While in the NIR and X-ray regime flares appear almost synchronously, a time delay of 1.5$\pm$0.5\,h between strong flares in the NIR and at 345\,GHz have been observed (\citealt{eckart2008a,eckart2008c}). A typical delay of $\sim$100\,min is assumed for high-energy flares to reach the sub-mm/mm regime (\citealt{meyer2008,marrone2008,yusef-zadeh2008}).
Since different wavelengths observe different areas of an opaque object, multi-frequency observations provide a radial view of the source. The traveling time that a NIR event needs to reach the 7\,mm area is the observed time delay between these observations and can therefore be used to connect flares in different frequencies to each other.\\
A cross correlation between NIR and 7\,mm data results in a time delay of (4.5$\pm$0.5)\,h with a correlation coefficient of 0.83. The error of 0.5\,h represents the range of coefficient values exceeding 68\%. This agrees well with the time delay of up to 5.25\,hours reported by \cite{yusef-zadeh2009} between NIR/X-ray and 43\,GHz observations.
If a jet feature were present, it is commonly accepted to have an appreciable inclination toward the line of sight (\citealt{markoff2007}). Therefore we adopt an inclination angle $\theta$ of $80^{\circ}$ for the following discussion. Lower inclinations are viable as well and will not significantly change the result. 
The observed radial separation of 1.5\,mas for a time delay of $(4.5\pm0.5)$\,h corresponds to a traveling velocity ($v_{t}$) of

$v_{t}=(0.4\pm0.2)\mathrm{mas\,h}^{-1}=(1.3\pm0.7)\cdot10^{8}\,\mathrm{m\,s}^{-1}=(0.4\pm0.3)\,c$.

Taking superluminal effects into account, the true gas velocity ($v_{g}$) can be calculated by

$v_{g}=\frac{v_{t}}{\sin{\theta}-\beta \cos{\theta}}=(0.4\pm0.3)\,c$.

The MHD model by \cite{yuan2009} proposes an expelled gas velocity ($v_{g}$) of 0.8\,c above the accretion disk shortly after the beginning of the flare, and therefore the presented velocity of $0.4$\,c can easily be achieved.
Even though there are other possible cooling processes of jets than adiabatic expansion, the time delay of ($4.5 \pm 0.5$)\,h agrees well with the current literature and is evidence for a causal connection between the NIR and radio events by this process. A correlation between NIR/X-ray and mm-flares together with a change in source morphology or size provides strong evidence for the adiabatic expansion model.

We have detected a mean closure phase of $(0.5\pm0.2)^{\circ}$ (LCP) and $(0.0\pm0.1)^{\circ}$ (RCP) for Sgr\,A* at 7\,mm using VLBI. While this mean value is zero within its error limits with no detections above $3\,\sigma$, there are a few detections with values exceeding the 2\,$\sigma$ range. For the closure triangle FD-KP-PT these high values coincide with the change of source structure on May\,17 8:00-10:00\,h UT (see Fig. \ref{sgra8-10}). Since the closure phases are a measure for the symmetry, this is another indication for the true existence of the change in the morphology of Sgr\,A*. This trend is not reproduced by the other triangles, but the observed and simulated values show that the highest deviations from zero-closure phases occur for triangles with an angle of maximum resolution close to the position angle of the detected secondary component.

\cite{fish2011} detected closure phase values of ($0\pm40$)$^{\circ}$ for Sgr\,A* using mm-VLBI at 1.3\,mm with an Earth-scaled triangle of SMT-JCMT-CARMA. \cite{broderick2011} showed that accretion flow models can reproduce closure phase values up to $\pm30^{\circ}$ in some cases. We report much lower closure phases at 7\,mm of $\approx 5^{\circ}$ , which are not excluded by theoretically expected values of (45-90)$^{\circ}$ at 1.3\,mm (\citealt{broderick2011}).

According to \cite{bower2014}, scattering still plays a significant role at 7\,mm. In this context a cause for a secondary component might be refractive noise, which can introduce artificial compact features on long baselines (\citealt{gwinn2014,johnson2015}). The refractive timescale for Sgr\,A* at 7\,mm is reported to be $\text{a}$\text{bout}\,three months (\citealt{akiyama2013}), which is much longer than the presented two-hour sub-images. We have only detected Sgr\,A* on May 17 at 7\,mm on baselines of up to 110 Mega-$\lambda$. The detection of a secondary component on shorter timescales and baselines as would be required for interstellar scintillation exclude this effect as a cause for the observed source structure.

In the case of an orbiting hot spot the deviation of the closure phases from zero depends on all properties affecting its symmetry and periodicity, such as the hot spot orbital size, its inclination, and the flux ratio between disk and hot spot. We have detected a change of morphology at a distance of 1.5\,mas from the emission center of Sgr\,A* (see Fig. \ref{sgra2hmaps}). \cite{reid2008} were able to rule out hot spots at radii above $\sim$80\,$\mu$as that contribute more than 30\% of the total 7\,mm flux. This limit also applies for variable adiabatically expanding off-center components. Even though fainter hot spots at larger radii may occur, the detected radial separation of 1.5\,mas on May 17 is much higher, and therefore a flux component of 0.02\,Jy at this scale cannot be related to a hot spot. A jet feature is not excluded by such a change in morphology. Since closure phases and 7\,mm maps provide independent tests for the source morphology, the presented consistent picture is a good indication for the correct detection of a secondary component.

\section{Summary and conclusions}
\label{conclusions}
This work presented observations of Sgr\,A* acquired in the framework of a three-day global observing campaign. A preceding NIR flare observed at the VLT triggered 7\,mm VLBA observations. Analysis of the DFT and mapped fluxes of Sgr\,A* showed a radio flare following the NIR observations by $\sim$(4.5$\pm$0.5)\,h on May 17, 2012 (9:30$\pm1$\,h UT). Splitting the data of this date into two-hour bins provided evidence of a possible change in source morphology shortly before the peak of the flare (8:00-10:00\,h UT). This source morphology was adequately modeled by two circular components well above the scattering size. The best-fitting model inherits a central component of 1.55\,Jy and a secondary 0.02\,Jy component separated by 1.5\,mas at $140^\circ$ (E-N). This two-component fit is superior to a single-component model for the specific time frame. 

This change in morphology should be detectable by non-zero closure phases because of its asymmetry. The phases of FD-KP-PT show increasing values during 8:00-10:00h UT with no values above 3\,$\sigma$, but some are in the 2\,$\sigma$ range. The other triangles show less prominent trends, but also have a few phases exceeding 2\,$\sigma$. The mean value of all individual closure phases as well as the total mean closure phase (0.5$\pm$0.2)$^{\circ}$ are close to zero for all times and all triangles within their errors. While the effect on closure phases is small and might be hidden within the error limits, we have tried to place constraints on this evidence by simulating several single- and two-component datasets using the Caltech VLBI analysis software \textsc{FAKE} and trying to reproduce the observed values. These simulations are strongly dependent on the seed chosen to generate random noise and can differ by $\sim$10$^\circ$ degree depending on this parameter. The observed effect on the closure phases is in the order of $\sim$5$^\circ$ and can therefore not be reliably tested by this method. But it can be gained from these simulations that all simulated closure phases are, like the observed values, zero within their error limits for the presented parameters. 

Furthermore, it is expected that a secondary component at a position angle of $140^\circ$ should have the strongest impact on closure triangles with an angle of maximum resolution close to this angle. This is true for all simulations, while depending on the applied random errors, it can change within any of the three triangles FD-KP-LA, FD-KP-PT, and FD-LA-PT. While KP-LA-PT is never the highest, for some cases it can offer the second highest mean closure phase. This is consistent with the presented observations, which show the strongest deviation from zero for FD-KP-PT on May 17. We can therefore not place entirely reliable constraints on this hypothesis, but we showed that it is possible to simulate datasets with a two-component model that can reproduce the observed trends. 

A secondary component present at the observed radial separation of $\sim$1.5\,mas would rule out hot spots, which can only occur on smaller radii, but it could be produced by an adiabatically expanding feature. The event would need traveling speeds of $\sim$0.4\,c to reach 1.5\,mas within ($4.5\pm0.5$)\,h, which is easily achievable according to the current literature.

\cite{bower2014} reported major-axis sizes of Sgr\,A* as an elliptical Gaussian of $35.4\times12.6$\,$R_{S}$ at a position angle of 95$^\circ$ east of north. This is much lower than the discussed source morphology due to a secondary component of 0.02\,Jy at 150\,$R_{S}$ at 140$^\circ$ east of north. 

We see indications of the existence of a secondary feature leading to an asymmetric morphology during the flaring state of Sgr\,A*. We cannot discard that these observed effects may be affected by random observational or weather effects. The observed morphology suggests an adiabatically expanding jet. Further observations are needed to clarify its existence. 

\small
\textit{Acknowledgments.} We would like to thank the anonymous referee for the constructive comments that further improved this paper.\\
The National Radio Astronomy Observatory is a facility of the National Science Foundation operated under cooperative agreement by Associated Universities, Inc..\\ 
C.R. was supported for this research through a scholarship from the International Max Planck Research School (IMPRS) for Astronomy and Astrophysics at the Universities of Bonn and Cologne.\\
E. R. was partially supported by the Spanish MINECO project AYA2012-38491-C02-01 and by the Generalitat Valenciana project PROMETEOII/2014/057, as well as by the COST MP0905 action ‘Black Holes in a Violent Universe’.

\end{document}